\documentclass[onecolumn]{openjournal}
\usepackage{graphicx}
\usepackage{amsmath}
\usepackage{amssymb}

\begin{document}

\title{Taking the long view: Rossby-number transition
and Rieger-type stellar cycles for 15,000 Kepler stars}

\author{Lavi Somers}
\email{lavisomers@gmail.com}
\affiliation{Physics Department, Technion -- Israel Institute of Technology,
Haifa 32000, Israel}

\author{Ilay Kamai}
\email{ilay.kamai@campus.technion.ac.il}
\affiliation{Physics Department, Technion -- Israel Institute of Technology,
Haifa 32000, Israel}

\author{Hagai B. Perets}
\email{hperets@physics.technion.ac.il}
\affiliation{Physics Department, Technion -- Israel Institute of Technology,
Haifa 32000, Israel}

\begin{abstract}

Stellar magnetic variability spans timescales from rotation, $P_{\mathrm{rot}}$,  to long-term activity cycles, $P_{\mathrm{cyc}}$. The Sun also exhibits the Rieger periodicity on intermediate timescales, but analogous variability in other stars remains poorly characterized. We develop a dedicated pipeline to measure Rieger-type intermediate periodicities in \emph{Kepler} photometry. We construct, to our knowledge, the largest catalog of individually measured stellar Rieger-type periods to date, measuring $P_{\mathrm{Rieg}}$ for 15,319 unique stars, several orders of magnitude more individual measurements than previously available, and use it to investigate the population-level behavior of stellar Rieger-type variability. The $P_{\mathrm{Rieg}}$--$P_{\mathrm{rot}}$ plane reveals four distinct populations: slow and intermediate rotators, a fast/low-$P_{\mathrm{Rieg}}$ population associated with $\gamma$~Doradus stars, and a fast/high-$P_{\mathrm{Rieg}}$ population dominated by tidally synchronized binaries. The intermediate regime shows a strong increase of $P_{\mathrm{Rieg}}$ with $P_{\mathrm{rot}}$, whereas this dependence weakens markedly in the slow-rotator regime, where $P_{\mathrm{Rieg}}$ clusters around $\sim150$ d, close to the canonical solar Rieger period of $\sim154$ d. Among late-type stars, this transition occurs at $\mathrm{Ro}_{\mathrm{tr}}=0.8\pm0.1$. The transition is more naturally organized by Rossby number than by rotation period alone, indicating a change in stellar magnetic behavior governed by the interaction between rotation and convection. The coincidence of this transition with independently reported changes in stellar magnetic behavior near $\mathrm{Ro}\sim1$ supports its interpretation as a change in dynamo configuration. The distinct tidally synchronized binary population provides a complementary test of this interpretation. These stars retain $P_{\mathrm{Rieg}}$ values characteristic of the slow population despite their anomalously rapid surface rotation and low inferred Rossby numbers, suggesting that tidal synchronization strongly modifies the envelope rotation while the deeper magnetic state traced by the Rieger-type periodicity remains closer to that of normal slow rotators. Together, these results suggest that Rieger-type variability traces a deeper global magnetic--rotational state, with the observed change in its relation to rotation near $\mathrm{Ro}\sim0.8$ reflecting a Rossby-number-dependent reorganization of the internal rotation--convection dynamo state.

\end{abstract}


\maketitle

\section{Introduction}
\label{sec:intro}

Stellar magnetic activity is observed over a broad hierarchy of timescales. At short timescales, stellar rotation produces quasi-periodic photometric modulation as active regions rotate across the visible hemisphere. On much longer timescales, the emergence and disappearance of active regions can vary systematically as part of a stellar magnetic-activity cycle, analogous to the approximately 11-year solar cycle. Between these two characteristic timescales, the Sun exhibits additional quasi-periodic variations on timescales of several months, the best known of which is the Rieger periodicity. In this work, we focus on this intermediate-timescale variability and its relation to stellar rotation.

Among these quantities, stellar rotation has been investigated most extensively. Rotation periods can be measured photometrically from the modulation produced by starspots and other active regions, and the continuous, high-precision photometry provided by \emph{Kepler} has enabled population-scale measurements for tens of thousands of stars. Several independent rotation-period catalogs have now been constructed, including samples containing approximately 34,030 stars in \citet{McQuillan2014}, approximately 39,592 stars in \citet{Santos2021}, and still larger samples in more recent analyses \citep[e.g.,][]{Reinhold2023,Kamai2026}. These catalogs have enabled extensive investigations of the dependence of stellar rotation on mass, temperature, age, activity, and evolutionary state ( \citealt{McQuillan2014,Santos2021,Mathur2023,Mathur2025}). Consequently, $P_{\mathrm{rot}}$ is one of the best-characterized observational quantities relevant to stellar magnetic activity in the \emph{Kepler} population.

Long-term magnetic activity cycles, $P_{\mathrm{cyc}}$, are also well established, although they are substantially less characterized at the population level than stellar rotation. The approximately 11-year solar cycle provides the canonical example, while the long-running Mount Wilson Ca~{\sc ii} H\&K survey established cyclic chromospheric variability in other late-type stars \citep{Wilson1978,Baliunas1995}. More recently, \citet{Reinhold2017} used four years of \emph{Kepler} photometry to identify photometric activity-cycle candidates in 3,203 of 23,601 stars, with periods between approximately 0.5 and 6~yr. The limited \emph{Kepler} baseline restricts robust measurements of longer, solar-like cycles, but these results demonstrate that long-timescale activity-related variability can nevertheless be identified photometrically in a substantial stellar sample.

In contrast to both rotation and long-term activity cycles, considerably less is known about magnetic variability on intermediate timescales. The canonical example is the solar Rieger periodicity, originally discovered as an approximately 154-day periodicity in $\gamma$-ray flares during solar cycle~21 \citep{Rieger1984}. Subsequent studies identified comparable periodicities in several independent indicators of solar magnetic activity, including X-ray and microwave flares, sunspot numbers and areas, radio emission, and magnetic-flux measurements \citep[e.g.,][]{Lean1989,Carbonell1990,Oliver1998,Gurgenashvili2016, Gurgenashvili2021}. Other periodicities at shorter and longer nearby timescales have also been reported and are commonly grouped under the broader designation of Rieger-type periodicities.

The solar Rieger periodicity is not a stationary clock. It is particularly prominent around maxima of solar activity, may persist for only part of an activity cycle, and its characteristic period changes with time. Long-term analyses of sunspot and magnetic-activity records have shown that Rieger power is intermittent and can vary substantially between solar cycles \citep[e.g.,][]{Lean1990,Oliver1998,Gurgenashvili2016}. \citet{Gurgenashvili2016}, for example, investigated solar cycles 14--24 and found Rieger periods of approximately $155$--$165$~days during stronger solar cycles and longer periods of approximately $185$--$195$~days during weaker cycles. They further found an anticorrelation between Rieger period and cycle strength. The dependence of the Rieger period on the evolving solar cycle suggests that this intermediate timescale contains information about the changing magnetic state of the Sun rather than representing a simple fixed harmonic of the rotation period.

The physical origin of the Rieger periodicity remains uncertain. A variety of mechanisms have been proposed, including periodic storage or emergence of magnetic flux, oblique-rotator or oscillator models, hydrodynamic Rossby waves, and magnetically modified Rossby waves \citep[e.g.,][]{BaiSturrock1991,Lou2000,Zaqarashvili2010,Gurgenashvili2016}. For the Sun, models involving magnetic Rossby waves in or near the tachocline have received particular attention because their characteristic frequencies can depend on rotation, differential rotation, and the internal magnetic field. In these models, unstable magnetic Rossby modes may modulate the emergence of magnetic flux and thereby produce periodic variations in surface magnetic activity. The dependence of the mode frequencies on the underlying toroidal magnetic field provides one possible explanation for the variation of the Rieger period between solar cycles \citep{Zaqarashvili2010,Gurgenashvili2016}. Under this solar interpretation, Rieger variability may provide an observable connection between surface magnetic activity and the dynamics of deeper layers, although whether the same mechanism operates across different stellar structures remains an open question.

Extending Rieger-type studies from the Sun to other stars provides an important opportunity to test these ideas. While observations of the Sun sample essentially a single stellar mass, rotation rate, and internal structure, stellar populations span a broad range of rotation periods, effective temperatures, convection-zone properties, convective turnover times, and magnetic-activity levels. If Rieger-type periodicities are controlled by rotationally influenced dynamo processes or global Rossby-like modes, systematic relations between $P_{\mathrm{Rieg}}$, $P_{\mathrm{rot}}$, and stellar properties should emerge across a sufficiently large population. In particular, the Rossby number, which combines the rotation period with the convective turnover time, provides a more physically motivated dynamo coordinate than rotation period alone and allows stars with different internal structures to be compared.

Observational studies of stellar Rieger-type periodicities remain sparse compared with the extensive literature on stellar rotation and long-term activity cycles. Short secondary activity periods have been reported in individual stars and small stellar samples, including signatures interpreted as possible Rieger-type variability \citep[e.g.,][]{Distefano2017,Gurgenashvili2022,Velloso2020}. However, these studies generally involve limited samples and heterogeneous diagnostics, and the physical origin of the detected intermediate-period signals is often ambiguous. As a result, it remains unclear whether stellar Rieger-type variability follows a coherent population-level relation with rotation and stellar properties, or instead reflects a mixture of distinct physical processes.

A more systematic stellar investigation was performed by \citet{Arkhypov2021}, who analyzed 1,726 selected \emph{Kepler} main-sequence stars. They measured characteristic activity periods from the temporal variation of the rotational modulation and associated Rieger-type periodicities with the shortest component, or lower boundary, of the resulting activity-period distribution. They found that the relation between these short periods and stellar rotation changes with effective temperature: the shortest periods were approximately independent of $P_{\mathrm{rot}}$ among cooler stars, whereas they approached an approximately proportional relation with $P_{\mathrm{rot}}$ among hotter stars. These regimes were interpreted in terms of possible Kelvin- and Rossby-wave drivers. Importantly, however, their Rieger-type relation was inferred statistically from the lower envelope of a broader distribution of activity periods rather than from a large catalog of individually selected intermediate-period measurements.

The intermediate Rieger-type regime therefore remains much less characterized than stellar rotation. Importantly, Rieger-type periodicities occur on timescales substantially shorter than long-term activity cycles, typically allowing several repetitions of the signal to be observed within the approximately four-year \emph{Kepler} baseline. This shorter timescale provides a major observational advantage over cycle-like variability and makes it possible to investigate intermediate magnetic variability across large stellar samples. Several fundamental questions nevertheless remain open: how $P_{\mathrm{Rieg}}$ scales with rotation across the main sequence, whether this scaling changes between stellar populations, and which stellar parameter provides the most natural coordinate for these changes. 

Beyond its relevance to stellar magnetic dynamics, understanding intermediate-timescale stellar variability may also be important for the analysis and interpretation of photometric light curves, particularly in the search for transiting exoplanets.

In this work, we use \emph{Kepler} photometry to investigate intermediate-timescale magnetic variability and its relation to stellar rotation. We adopt rotation periods from several independent catalogs containing tens of thousands of stars and develop a dedicated procedure to search for periodicities on timescales longer than rotation while detrending long-timescale variability. We then test whether the resulting intermediate-period signals are consistent with a common Rieger-type interpretation. We measure $P_{\mathrm{Rieg}}$ for 15,319 unique \emph{Kepler} stars, to our knowledge the largest sample assembled for this phenomenon. This large sample enables us to investigate, for the first time at population scale, how Rieger-type periodicities are organized with respect to stellar rotation. The resulting $P_{\mathrm{Rieg}}$--$P_{\mathrm{rot}}$ distribution reveals distinct stellar populations and scaling regimes. In particular, the slow-rotator population has a characteristic $P_{\mathrm{Rieg}}\sim150$~d, placing the canonical $\sim154$-day solar Rieger periodicity naturally within this stellar regime. We also identify a population of anomalously rapid rotators associated with synchronized binaries. Despite their rapid surface rotation, these stars retain $P_{\mathrm{Rieg}}$ values characteristic of the slow-rotator population, suggesting that tidal synchronization modifies the observed rotation without producing a corresponding shift in the magnetic state traced by the Rieger-type periodicity. We further investigate whether the transitions between the rotational regimes are more naturally organized by rotation period, effective temperature, Rossby number, or other stellar properties, and whether Rieger-type variability is tied primarily to the observed surface rotation or instead traces a deeper magnetic state whose coupling to rotation changes across dynamo regimes.
\section{Data and Methods}
\label{sec:data}

\subsection{Kepler Data, Rotation Periods, and Stellar Properties}
\label{sec:stellar_properties}

We measured Rieger-type periodicities from \emph{Kepler} light curves, adopting stellar rotation periods from published catalogs. We additionally measured long-term photometric activity-cycle periods for use as an independent magnetic-variability reference sample. We used the long-cadence light curves from \emph{Kepler} Data Release~25 \citep{kepler_dr25}, processed for instrumental systematics with version~9.3 of the PDC-MAP pipeline. For each target, we included all available observing quarters, up to 18 quarters in total, at a cadence of approximately 30 minutes. The light curves are publicly available through MAST\footnote{\url{https://doi.org/10.17909/T9488N}}.

We adopted stellar rotation periods, $P_{\mathrm{rot}}$, from four independent \emph{Kepler} catalogs: \citet{McQuillan2014}, containing 34,030 stars; \citet{Santos2021}, containing approximately 39,592 stars; \citet{Reinhold2023}, containing approximately 67,515 stars; and the Maunder catalog of \citet{Kamai2026}, containing approximately 148,746 stars. For the Maunder catalog, we retained only the 119,428 stars with CI80 smaller than 0.4, corresponding to the high-confidence rotation-period sample.

We combined multiple catalogs because no single rotation-period catalog provides a definitive reference for the full \emph{Kepler} sample. The catalogs differ in their period-detection methods, sample-selection criteria, quality cuts, and treatment of aliases and harmonics, and therefore exhibit distinct strengths, limitations, and selection effects. Combining them increases the number of stars with available rotation-period measurements and enables us to test the robustness of our results against catalog-specific biases.

For validation of our long-term variability measurements, we used the activity-cycle catalog of \citet{Reinhold2017}, containing 3,203 \emph{Kepler} stars.

We used the catalog of \citet{GodoyRivera2025} to identify binary systems and as the basis for the Rossby-number calculations adopted in this work. Stellar effective temperatures, surface gravities, metallicities, masses, and radii were taken from the \emph{Kepler} Q1-Q17 Data Release~25 Stellar Properties Catalog \citep{Mathur2017}.

\subsection{Measurement of Long-Term Activity-Cycle Periods}
\label{sec:data cycle}

As an approximate reference for the long-term variability timescale, we also extracted photometric activity-cycle periods, $P_{\mathrm{cyc}}$, using a pipeline based on the method of \citet{Reinhold2017}. While \citet{Reinhold2017} applied their analysis to stars from the \citet{McQuillan2014} rotation catalog, we applied the same procedure independently to each of the four rotation-period catalogs adopted in this work. When restricted to the \citet{McQuillan2014} sample, our recovered $P_{\mathrm{cyc}}$ values show close agreement with those reported by \citet{Reinhold2017}, providing a validation of our implementation. For the present work, we adopted a more permissive FAP threshold of $10\%$ and an additional variability criterion of $R_{\mathrm{var,max}}>0.01$; see Section~\ref{sec:data rieger}. These measurements are not analyzed here as an independent activity-cycle sample. Instead, they are used only as a rough empirical estimate of the long-timescale variability domain and as a guide for separating the intermediate Rieger-type search from longer-term activity variations.

\subsection{Measurement of Rieger-Type Periods}
\label{sec:data rieger}

During our analysis of the stellar light curves, we identified a distinct periodicity in addition to the rotation and long-term activity-cycle periods. Based on its characteristic intermediate timescale, we interpret this signal as a Rieger-type periodicity. Its observational properties and astrophysical interpretation are examined in detail in Section~\ref{sec:rieger_results}.

\begin{figure}[t]
    \centering
    \includegraphics[width=1\linewidth]{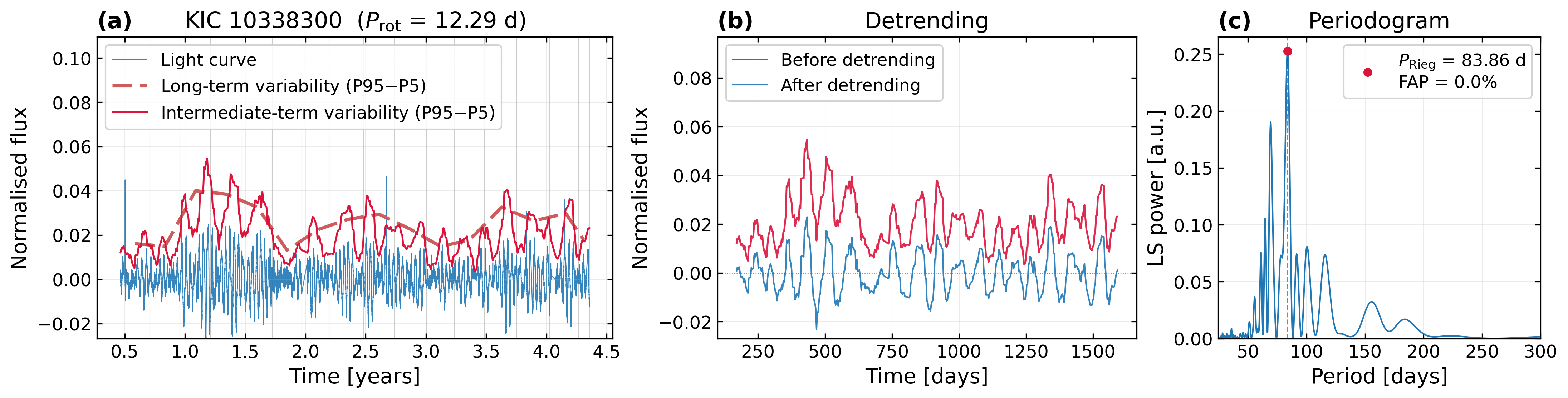}
    \caption{Example of the measurement of Rieger-type variability for KIC~10338300. Panel (a) shows the normalized flux together with the long-term and intermediate-term variability components. Panel (b) illustrates the effect of detrending in isolating the intermediate-term variability. Panel (c) shows the Lomb--Scargle periodogram of the detrended intermediate-term variability.}
    \label{fig:light-curve-rieger}
\end{figure}

We developed a dedicated pipeline to search for Rieger-type periodicities on timescales longer than the stellar rotation period and below the characteristic domain of long-term activity variability. For each available catalog measurement of $P_{\mathrm{rot}}$, the pipeline was applied independently; an example is shown in Figure~\ref{fig:light-curve-rieger}.

Each observing quarter was normalized by its median flux and shifted to a median value of unity. Residual long-term instrumental trends within each quarter were removed by subtracting a second-order polynomial fit. We then divided each light curve into overlapping windows of duration $P_{\mathrm{rot}}$. Adjacent windows overlapped by 75\%, corresponding to a time step of $0.25P_{\mathrm{rot}}$.

For each window centered at time $t_j$, we calculated the photometric variability range
$$
R_{\mathrm{var}}(t_j) = F_{95}(t_j)-F_{5}(t_j),
$$

where $F_{95}(t_j)$ and $F_{5}(t_j)$ are the 95th and 5th percentiles, respectively, of the normalized flux within the window. The resulting local-variability time series traces temporal changes in the amplitude of the rotational modulation. 

We also recorded the maximum measured variability,
$$
R_{\mathrm{var,max}} = \max_j\left[R_{\mathrm{var}}(t_j)\right].
$$

The $R_{\mathrm{var}}(t)$ time series may contain both intermediate-timescale variability and a slowly varying long-timescale component. We removed the slowly varying component using a running-mean subtraction filter,
$$
\widetilde{R}_{\mathrm{var},i}(t;W) = R_{\mathrm{var},i}(t) - \left\langle R_{\mathrm{var},i}(t)\right\rangle_W,
$$
where $W$ is the width of the running-mean window. The optimal value of $W$ is not known \emph{a priori}: the window must be sufficiently short to suppress long-term variations while remaining sufficiently long to preserve the unknown Rieger-type signal. Because the period domain transfer function of the filter depends strongly on $W$, the recovered period may be biased when it approaches the filter scale. To reduce the sensitivity of the measurement to the specific choice of the detrending window, we sampled 50 values of $W$, linearly spaced over the range
$$
2P_{\mathrm{rot}}
\leq W
\leq
450~\mathrm{d}
$$

We did not apply additional higher-order detrending procedures, such as local variance normalization of the $R_{\mathrm{var}}$ time series. Such transformations may introduce artificial variability or suppress physically meaningful information contained in the temporal evolution of the rotational-modulation amplitude.

\begin{figure}[t]
    \centering
    \includegraphics[width=0.75\linewidth]{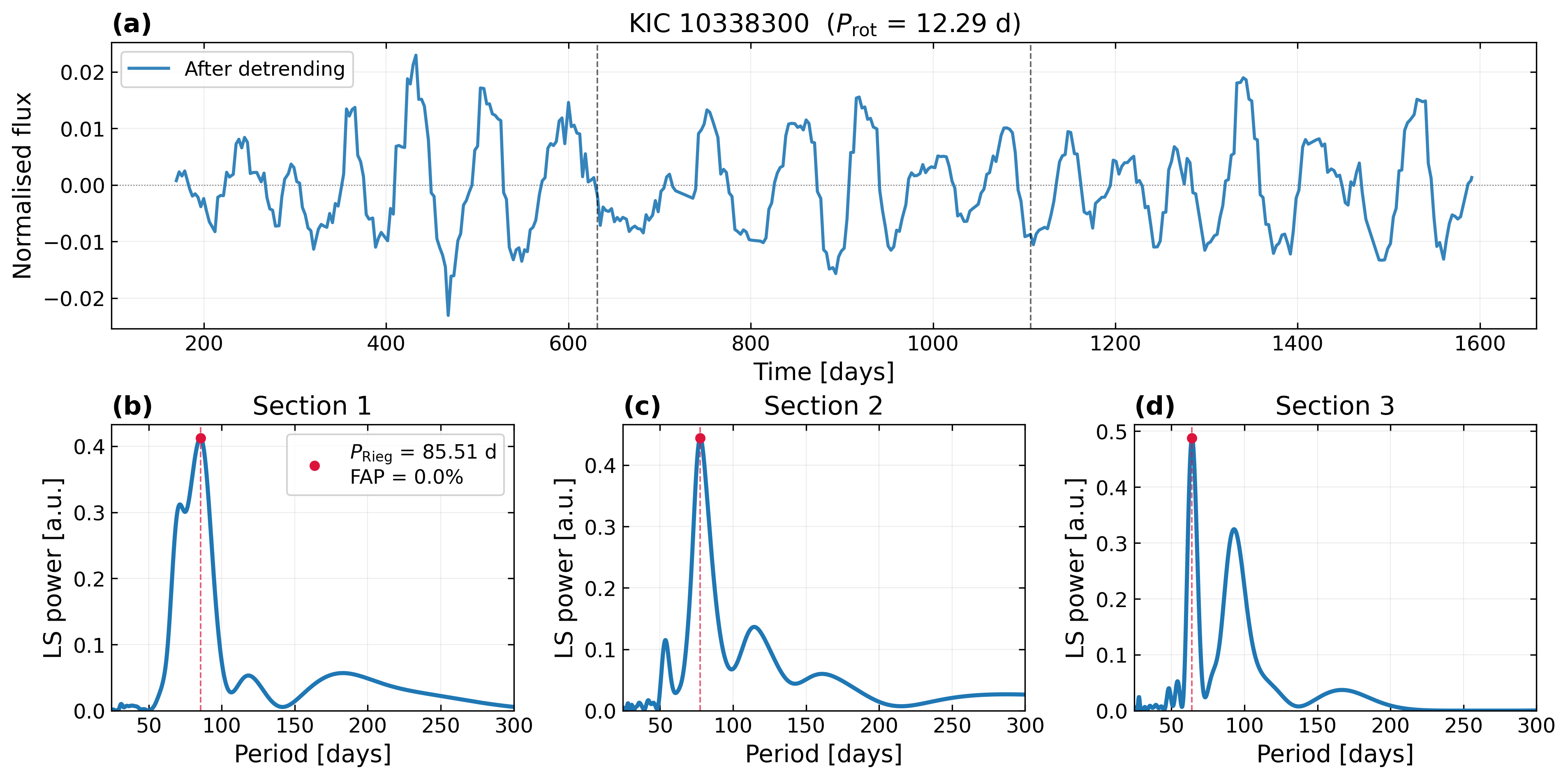}
    \caption{Example of the Rieger-type periodicity measurement for KIC~10338300. Panel (a) shows the light curve divided into three temporal segments. The corresponding periodograms in panels (b)--(d) yield Rieger-type periods of 85.5, 77.5, and 64.0 days, respectively, illustrating the temporal variability of the Rieger-type signal. The final adopted period, calculated from the two closest measurements, is 81.5 days.}
    \label{fig:light-curve-rieger-3-sections}
\end{figure}

Because Rieger-type periodicities may vary in period, amplitude, or coherence over the full \emph{Kepler} observing baseline, we divided each $R_{\mathrm{var}}(t)$ time series into three equal-duration segments of approximately 1.4 years, as illustrated in Figure~\ref{fig:light-curve-rieger-3-sections}. Each segment, denoted $R_{\mathrm{var},s}(t)$, was analyzed independently. This segmentation mitigates the effects of temporal evolution and improves the sensitivity to intermittent or weakly coherent Rieger-type signals.

For each temporal segment, indexed by $s$, and each detrending-window width, $W$, we calculated a Lomb--Scargle periodogram of $\widetilde{R}_{\mathrm{var},s}(t;W)$ and searched for candidate periods within the range
$$
2P_{\mathrm{rot}}
\leq P_{\mathrm{Rieg}}
\leq 300~\mathrm{d}.
$$
The lower limit of $2P_{\mathrm{rot}}$ separates the intermediate-timescale signal from the rotation period and its nearby harmonics and lies safely above the formal Nyquist limit associated with the $0.25P_{\mathrm{rot}}$ sampling interval. The upper limit of 300~days encompasses the expected range of stellar analogs of the solar $\sim154$-day Rieger periodicity while remaining broadly separated from the longer-term variability indicated by our $P_{\mathrm{cyc}}$ measurements. These $P_{\mathrm{cyc}}$ values were used only to guide the adopted upper limit and were not applied as star-by-star constraints.

Within this search range, we identified the highest Lomb--Scargle periodogram peak for each $(W,s)$ combination and denote its period and power by $P(W,s)$ and $H(W,s)$, respectively.

For each segment, the candidate period was defined as the period associated with the highest power obtained across all tested detrending-window widths:
$$
P_s=P(W_s^\ast,s),
\qquad
W_s^\ast=\operatorname*{arg\,max}_{W} H(W,s).
$$

Where $P(W, s)$ and $H(W, s)$ denote the period and power of the highest periodogram peak obtained using detrending-window width (W). 

We quantified the robustness of $P_s$ using a stability metric. The stability $S_s$ was defined as the fraction of tested detrending-window widths for which the tracked period $P(W,s)$ remained within 3\% of $P_s$.  

For some detrending-window widths, particularly those far from $W_s^\ast$, the highest periodogram peak can shift abruptly to a different period because a competing peak becomes stronger. Such exchanges in peak ranking can produce artificial discontinuities in the sequence of measured periods, even when the candidate signal remains present as the second- or third-highest peak. These changes may result from the window-dependent transfer function of the running-mean subtraction filter, which can attenuate the candidate peak relative to nearby peaks. 

\begin{figure}[t]
    \centering
    \includegraphics[width=0.5\linewidth]{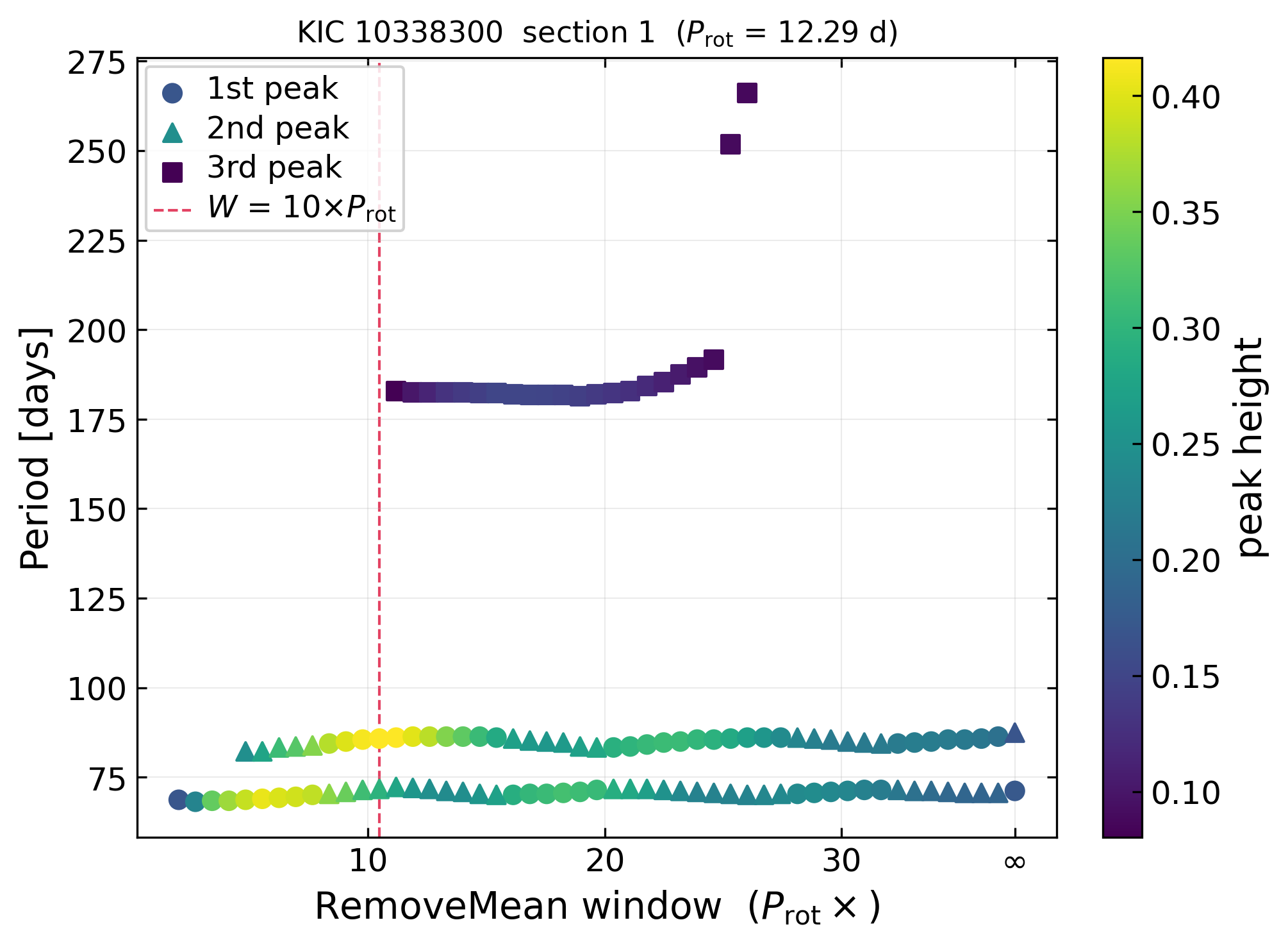}
    \caption{Stability metric calculation for the first temporal section ($s=1$) of KIC~10338300. Two stable, competing periodicities are evident, whereas the third varies strongly with the detrending-window width ($W$). The globally strongest periodogram peak occurs at $P=85.5$ days for $W = 10P_{\mathrm{rot}}$ where $P_{\mathrm{rot}}=12.29$ days, and has stability of 92\%. The competing peak at 69 days has stability of 100\% while the peak at 182 days has stability of 22\% }
    \label{fig:light-curve-rieger-window sweep}
\end{figure}

Because the stability metric is intended to measure gradual changes in the recovered period rather than discrete changes in peak ranking, we tracked the candidate signal using the three strongest periodogram peaks at each value of $W$. Whenever the highest peak shifted away from $P_s$, we selected, when available, the retained peak closest to $P_s$, as illustrated in Figure~\ref{fig:light-curve-rieger-window sweep}. 

The resulting stability metric therefore measures whether the candidate period persists continuously across detrending scales, largely independently of changes in its instantaneous periodogram rank. 

A genuine astrophysical periodicity is expected to remain detectable over a broad range of detrending-window widths, whereas a feature produced by a particular choice of $W$ is expected to exhibit low stability. In the analysis presented in Section~\ref{sec:rieger_results}, we retained candidates satisfying $S_s\geq40\%$.

We estimated the false-alarm probability (FAP) of each candidate period, \(P_s\), using the analytical approximation for the generalized Lomb--Scargle periodogram, $\mathrm{FAP} \approx M(1-p_0)^{(N-3)/2}$ where \(p_0\) is the standard-normalized peak power, \(N\) is the number of data points, and $M = T\left(\frac{1}{P_{\min}}-\frac{1}{P_{\max}}\right)$ is the estimated number of independent frequencies across the searched period range \citep{Horne1986,Zechmeister2009,VanderPlas2018}. This expression assumes independent and identically distributed Gaussian white noise. For a representative subset of targets, we compared the analytical FAP values with permutation-based estimates and found them to be broadly consistent. We therefore adopted the analytical calculation for the full sample, because permutation testing of the densely sampled $R_{\mathrm{var}}$ time series would be computationally prohibitive. In Section~\ref{sec:rieger_results}, we retained only candidates with $\mathrm{FAP}<0.01\%$.

Finally, we required each star to have at least 16 valid \emph{Kepler} quarters. Because Rieger-type periodicity is less well established in stellar observations, we chose to restrict the analysis to the highest-quality light curves available in order to minimize potential biases and systematic effects. We further required the candidate period in each of the three temporal segments to satisfy both the stability and FAP criteria. When all three segments passed these criteria, the final Rieger-type period, $P_{\mathrm{Rieg}}$, was defined as the arithmetic mean of the two most closely matching values among $P_1$, $P_2$, and $P_3$, where the subscripts here denote the three temporal segments. This procedure is motivated by the quasi-periodic nature of Rieger-type variability: the dominant timescale may evolve over the observing baseline rather than remain strictly stationary. We therefore use $P_{\mathrm{Rieg}}$ to denote a characteristic dominant timescale, estimated from the two most mutually consistent temporal measurements, rather than an invariant clock. This approach allows for moderate temporal evolution of the signal while reducing the influence of an anomalous period measurement in any single segment. 

\subsection{Rossby-Number Calculation}
\label{sec:Ro}

One of the key physical parameters governing stellar magnetic-dynamo behavior is the Rossby number, which quantifies the relative importance of rotation and convection and is defined as
$$
\mathrm{Ro}=\frac{P_{\mathrm{rot}}}{\tau_c},
$$
where $P_{\mathrm{rot}}$ is the stellar rotation period and $\tau_c$ is the convective turnover time. Rossby numbers were computed using the empirical $\tau_c(B-V)$ calibration of \citet{Noyes1984a}. Since homogeneous Johnson $(B-V)$ measurements are not available for the full sample, we used the dereddened Gaia colors $(BP-RP)_0$ reported by \citet{GodoyRivera2025} and converted them to $(B-V)$ using the dwarf color sequence of \citet{PecautMamajek2013}.

For quantitative analyses involving Rossby number, we restrict the sample to stars with a \citet{GodoyRivera2025} color measurement, converted $(B-V)$ within the calibrated range of the \citet{Noyes1984a} relation ($0.40\leq B-V\leq1.40$), and a main-sequence classification according to the $\log g$--$T_{\mathrm{eff}}$ criterion of \citet{Ciardi2011}. This calibrated subset comprises $77\%$ of the sample. For descriptive purposes, we also report formally extrapolated Rossby numbers outside the calibrated color interval where possible; these values are shown only for reference and are not used in the quantitative Rossby-number analysis.
\section{Results}
\label{sec:results}

We applied the Rieger-type period measurement procedure described in Section~\ref{sec:data} independently to each of the four adopted rotation-period catalogs.

We provide two versions of the resulting catalogs. The first contains the selected samples used in the scientific analysis presented below. The second contains the full set of measurements, together with the associated diagnostic quantities, including the false-alarm probabilities and stability metrics. The full catalogs therefore allow users to apply alternative quality criteria appropriate to their specific applications.

\subsection{Rieger-Type Period Measurements and Validation}
\label{sec:rieger_results}

We searched for Rieger-type periodicities on intermediate timescales longer than the stellar rotation period and shorter than the characteristic long-term activity timescale, following the procedure described in Section~\ref{sec:data rieger}. We consider a Rieger-type period, $P_{\mathrm{Rieg}}$, to be valid only when the corresponding periodogram peak satisfies both $\mathrm{FAP}_s<0.01\%$ and $S_s>40\%$ in each of the three temporal segments, $s=1,2,3$. We additionally require $R_{\mathrm{var,max}}\geq0.002$ to exclude the lowest-variability light curves, which are dominated by instrumental systematics.

Applying these selection criteria yields 12,055 stars in the \citet{McQuillan2014} catalog, 7,829 in \citet{Santos2021}, 12,344 in \citet{Reinhold2023}, and 13,471 in \citet{Kamai2026}. Combining the four rotation-period catalogs yields 15,319 unique stars with a measured Rieger-type period, of which 13,141 appear in at least two catalogs. Despite differences in the adopted $P_{\mathrm{rot}}$ values, the recovered $P_{\mathrm{Rieg}}$ is highly consistent across catalogs: for 78\% of these stars, the catalog-specific measurements agree to within 2\%, with a median fractional spread of only 0.5\%.

For the population-level analysis below, we therefore also define a pooled sample containing all 45,699 catalog-specific measurements from the four rotation catalogs. Because the same star may appear in more than one catalog, stars with measurements in multiple catalogs contribute more than once to the pooled sample and therefore receive greater effective weight. This is justified by the strong cross-catalog consistency of the recovered $P_{\mathrm{Rieg}}$ values described above. At the same time, stars appearing in only one rotation catalog are retained, allowing the pooled sample to incorporate the full set of catalog-unique detections.

In order to verify our method, we applied the same pipeline to the solar SATIRE-S photometric data \citep{Yeo2014} and recovered a variable Rieger-type period between approximately 130 and 200~d across different solar cycles. Overall, the periods recovered by our pipeline agree well with the solar measurements of \citet{Gurgenashvili2016} and \citet{Gurgenashvili2021}. Despite differences in the activity indicators and analysis methods, both our results and the previous solar analyses show that the Rieger-type period is not fixed, but varies from one solar cycle to another.

\begin{figure}[t]
    \centering
    \includegraphics[width=0.75\linewidth]{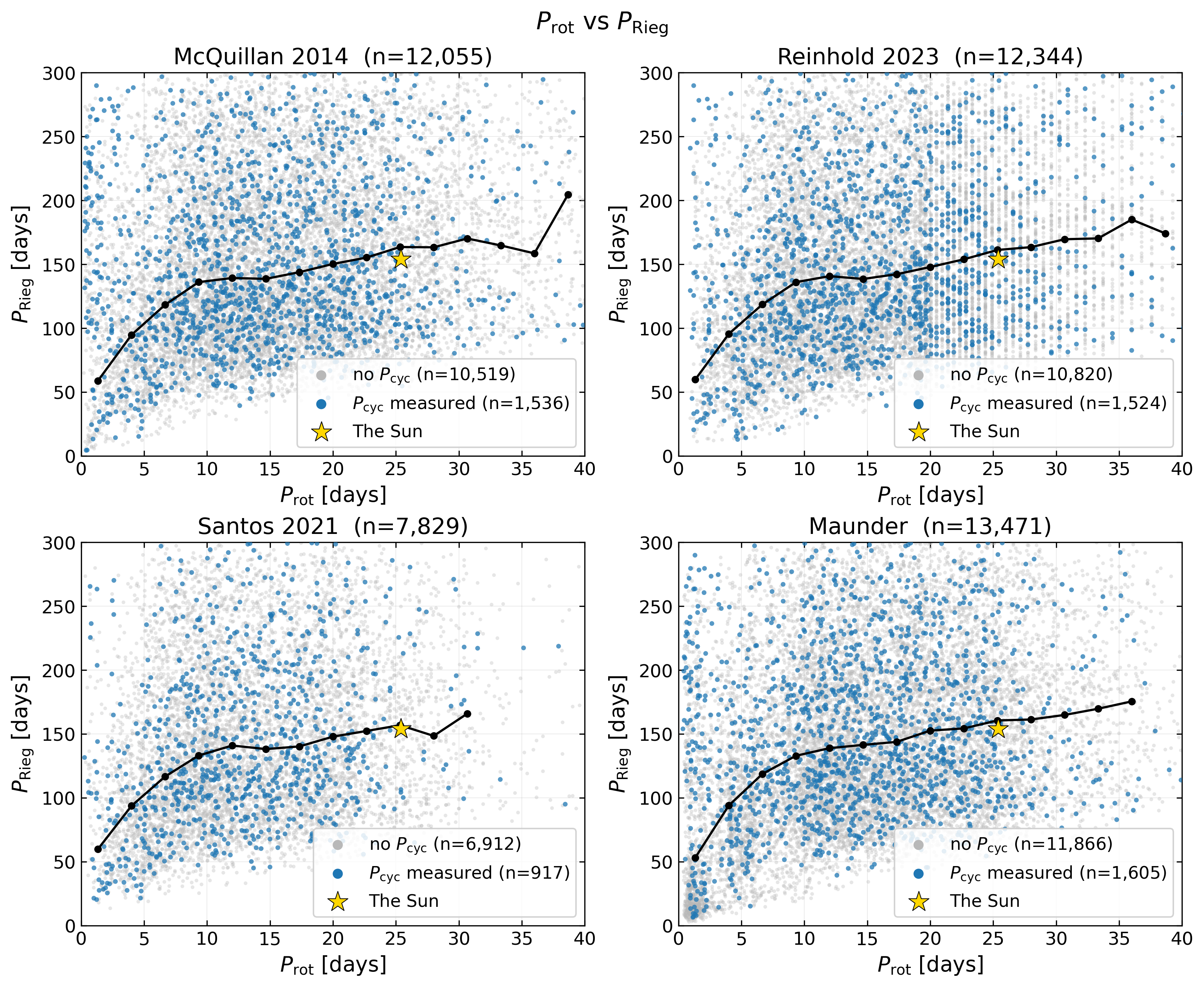}
    \caption{$P_{\mathrm{Rieg}}$ as a function of $P_{\mathrm{rot}}$; stars with measured $P_{\mathrm{cyc}}$ are highlighted in blue. The median $P_{\mathrm{Rieg}}$ values in 15 $P_{\mathrm{rot}}$ bins are overplotted as a black line.
}
    \label{fig:Prieg vs Prot color Pcyc}
\end{figure}

We first analyzed the relation between $P_{\mathrm{Rieg}}$ and $P_{\mathrm{rot}}$ for stars with a measured $P_{\mathrm{cyc}}$, separately for each of the four rotation-period catalogs considered in Section~\ref{sec:data}. We then applied the same analysis to stars without a measured $P_{\mathrm{cyc}}$. As shown in Figure~\ref{fig:Prieg vs Prot color Pcyc}, the two populations exhibit similar behavior in the $P_{\mathrm{Rieg}}$--$P_{\mathrm{rot}}$ plane. This consistency supports the interpretation that the intermediate periods identified by our pipeline are Rieger-type periodicities even when no direct measurement of $P_{\mathrm{cyc}}$ is available.

The number of stars with both valid $P_{\mathrm{cyc}}$ and $P_{\mathrm{Rieg}}$ measurements is relatively small. Importantly, the absence of a measured activity cycle does not imply the absence of long-term magnetic variability: the cycle period may lie outside the sensitivity range of the $P_{\mathrm{cyc}}$ measurement method for the \emph{Kepler} baseline, or the photometric variability may be insufficient for a reliable detection. Including stars without a measured $P_{\mathrm{cyc}}$ therefore increases the available $P_{\mathrm{Rieg}}$ sample by approximately a factor of seven.

Despite substantial differences in their original sizes, the four rotation-period catalogs yield comparable numbers of stars with valid $P_{\mathrm{Rieg}}$ measurements after application of our selection criteria. The minimum variability requirement, $R_{\mathrm{var,max}}\geq0.002$, first reduces each catalog to roughly 25,000 stars suitable for the Rieger-type search. Requiring a low $FAP$ and stable intermediate-timescale signal then reduces these samples to approximately 10,000 stars per catalog with valid $P_{\mathrm{Rieg}}$ detections. This convergence is driven primarily by the larger fraction of low-variability stars in the more extensive rotation-period catalogs, which are preferentially removed by the variability cut.

\subsection{Population Structure in the $P_{\mathrm{Rieg}}$--$P_{\mathrm{rot}}$ Plane}
\label{sec:RiegerRotationalRegimes}

\begin{figure}[t]
    \centering
    \includegraphics[width=0.75\linewidth]{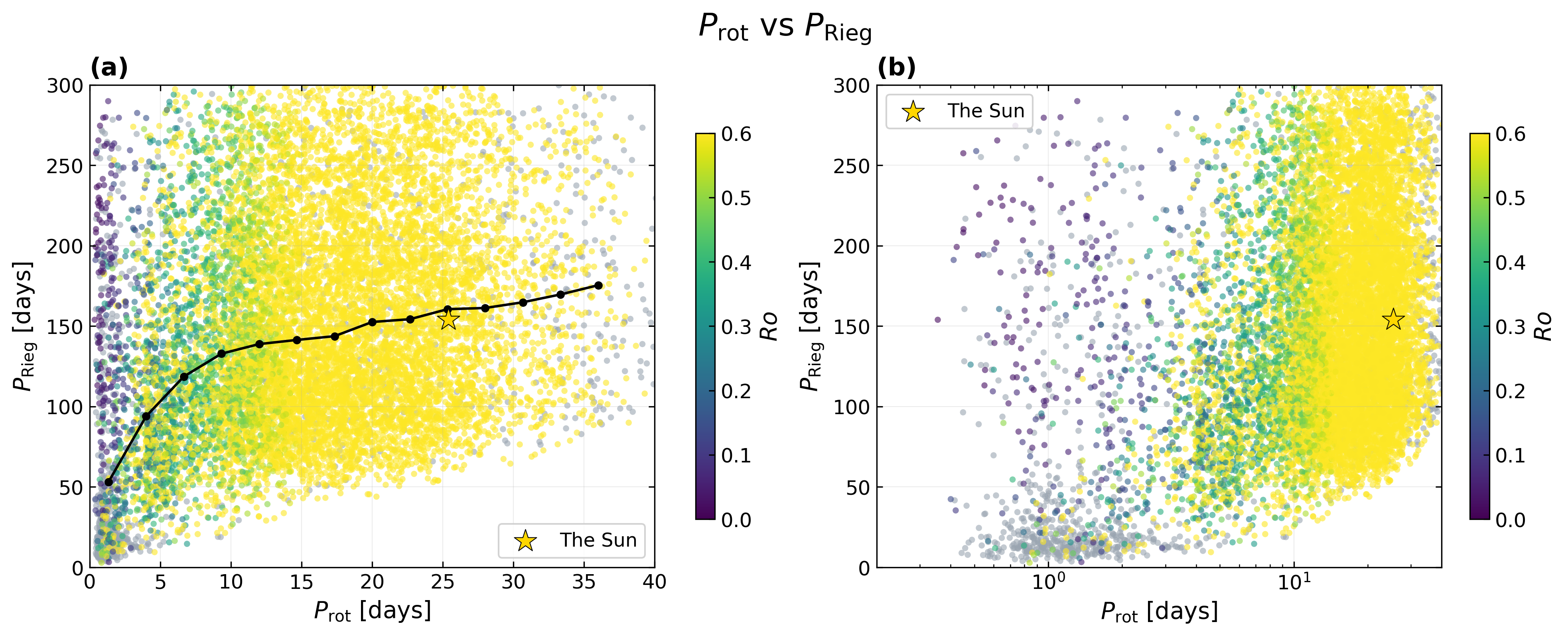}
    \caption{$P_{\mathrm{Rieg}}$ versus $P_{\mathrm{rot}}$ for the Maunder catalog, colored by Rossby number. Stars without an available Rossby number are shown in gray. The black curve indicates the binned median relation after excluding the anomalous population. Panel (a) shows the relation on a linear $P_{\mathrm{rot}}$ scale, while panel (b) shows the same data with a logarithmic $P_{\mathrm{rot}}$ axis.}
    \label{fig:Prieg vs Prot color Ro}
\end{figure}

Having established the robustness and consistency of the measured $P_{\mathrm{Rieg}}$ values, we next use the full sample to examine how Rieger-type periodicities are organized with respect to stellar rotation. The $P_{\mathrm{Rieg}}$--$P_{\mathrm{rot}}$ distribution shown in Figure~\ref{fig:Prieg vs Prot color Ro} is strongly non-linear and reveals four distinct populations. One of these consists of rapidly rotating stars with unusually long $P_{\mathrm{Rieg}}$, which we show in Section~\ref{sec:Binaries} is predominantly composed of binary systems; we therefore exclude this anomalous population from the following rotational-regime analysis. The remaining stars define three distinct fast, intermediate, and slow rotational regimes. As shown in Figure~\ref{fig:Prieg vs Prot color Pcyc}, the different rotation-period catalogs exhibit broadly similar $P_{\mathrm{Rieg}}$--$P_{\mathrm{rot}}$ behavior over their common parameter range. We therefore focus in Figure~\ref{fig:Prieg vs Prot color Ro} on the Maunder catalog, which is the only catalog that clearly resolves the low-$P_{\mathrm{Rieg}}$ fast-rotator population, owing to its substantially larger number of stars with $P_{\mathrm{rot}}<1$~d.

At fixed rotation period, the measured $P_{\mathrm{Rieg}}$ values show substantial scatter. Part of this spread may arise from observational effects, including photometric noise and differences in stellar inclination, which affect the visibility and evolution of surface magnetic structures. In addition, the Rieger-type timescale itself is not expected to remain strictly constant in time. As demonstrated by our SATIRE-S solar validation in Section~\ref{sec:rieger_results}, the recovered Rieger period varies between solar cycles, consistent with previous solar measurements. Thus, the observed spread at fixed $P_{\mathrm{rot}}$ likely reflects a combination of observational effects and intrinsic temporal variability of the Rieger-type clock.

\subsection{Slow, Intermediate, and Fast-Low-$P_{\mathrm{Rieg}}$ Regimes}
\label{sec:SlowInterFastRegimes}

\begin{figure}[t]
    \centering
    \includegraphics[width=0.75\linewidth]{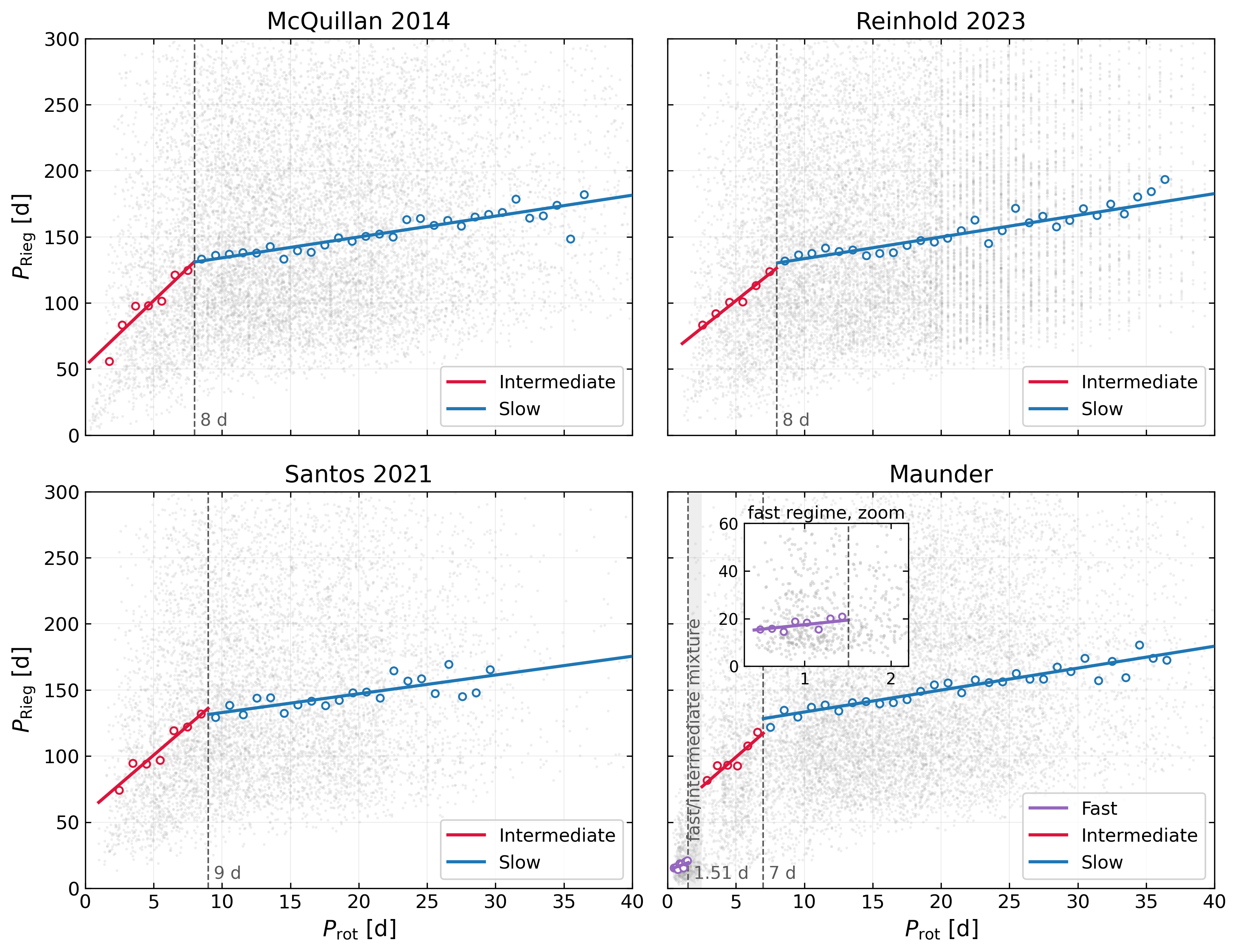}
    \caption{Binned $P_{\mathrm{Rieg}}$--$P_{\mathrm{rot}}$ relations and linear fits for the rotational regimes in each catalog. Dashed lines indicate the fitted regime boundaries, and the inset highlights the fast-rotator range. The anomalous population is excluded prior to fitting.
}
    \label{fig:prieg fit}
\end{figure}

Having identified the four populations morphologically, we first quantify the three non-anomalous rotational regimes after excluding the fast-high-$P_{\mathrm{Rieg}}$ population. To determine the locations of the transitions between the rotational regimes, we fitted the median-binned $P_{\mathrm{Rieg}}$--$P_{\mathrm{rot}}$ relation with a continuous piecewise-linear model of the form

$$
y = A + Bx + C\max(0,x-t_1) + D\max(0,x-t_2),
$$

where $y$ and $x$ represent $P_{\mathrm{Rieg}}$ and $P_{\mathrm{rot}}$, respectively, or $\log P_{\mathrm{Rieg}}$ and $\log P_{\mathrm{rot}}$ for the fast--intermediate transition, and $t_1$ and $t_2$ denote the locations of changes in slope. The transition locations were determined by sweeping over $t_1$ and $t_2$ and minimizing the fit residuals. For the intermediate--slow transition, only a single break is apparent, and we therefore set $D=0$. For the fast--intermediate transition, both break terms were retained because the two populations overlap over a finite range in $P_{\mathrm{rot}}$, producing an intermediate region with a different effective slope. After determining the regime boundaries, we refitted each regime separately with a simple linear relation,

$$
y=A+Bx,
$$

in order to characterize the $P_{\mathrm{Rieg}}$--$P_{\mathrm{rot}}$ relation within each regime. The resulting fits and inferred regime boundaries are shown in Figure~\ref{fig:prieg fit}.

The intermediate rotational regime exhibits the strongest dependence of $P_{\mathrm{Rieg}}$ on $P_{\mathrm{rot}}$, producing a well-defined rising sequence. The transition to the slow rotational regime occurs at $P_{\mathrm{rot}}\simeq7$--$9$~d, depending on the rotation-period catalog. Beyond this transition, the dependence of $P_{\mathrm{Rieg}}$ on $P_{\mathrm{rot}}$ becomes considerably weaker. As quantified in Table~\ref{tab:regime_ab}, the fitted slope decreases from approximately $9~\mathrm{d,d^{-1}}$ in the intermediate regime to only $\sim1.5~\mathrm{d,d^{-1}}$ in the slow regime. The slow-rotator relation is therefore characterized by only a weak increase in $P_{\mathrm{Rieg}}$ with $P_{\mathrm{rot}}$, remaining clustered around $P_{\mathrm{Rieg}}\simeq150$~d rather than showing a strong rotational scaling, in excellent agreement with the canonical solar Rieger periodicity of approximately 154~d.

A third rotational regime appears at the shortest rotation periods and is resolved only in the Maunder catalog. The fitted fast-regime boundary occurs at $P_{\mathrm{rot}}\simeq1.5$~d. The transition is not sharp: between approximately $1.5$ and $3.5$~d, the fast and intermediate populations overlap. This overlap is also why the intermediate-regime fit in Table~\ref{tab:regime_ab} is restricted to $P_{\mathrm{rot}}>2.5$~d. At $P_{\mathrm{rot}}\lesssim1.5$~d, the fast population shows only a weak dependence of $P_{\mathrm{Rieg}}$ on $P_{\mathrm{rot}}$, with a fitted slope consistent with zero within the uncertainties and a nearly constant $P_{\mathrm{Rieg}}$ of approximately 14~days.

\begin{table*}[t]
    \centering
    \begin{tabular}{l|cc|cc|cc}
    & \multicolumn{2}{c|}{Fast} & \multicolumn{2}{c|}{Intermediate} & \multicolumn{2}{c}{Slow} \\
    Catalog & $a$ [d~d$^{-1}$] & $b$ [d] & $a$ [d~d$^{-1}$] & $b$ [d] & $a$ [d~d$^{-1}$] & $b$ [d] \\
    \hline
    McQuillan 2014 & $--$           & $--$      & $9.51\pm1.09$    & $54.6\pm6.2$  & $1.58\pm0.15$  & $118.3\pm3.0$ \\
    Reinhold 2023  & $--$           & $--$      & $8.01\pm1.16$    & $62.1\pm6.6$  & $1.64\pm0.14$  & $117.1\pm2.7$ \\
    Santos 2021    & $--$           & $--$      & $9.06\pm0.86$    & $55\pm5.2$  & $1.42\pm0.25$  & $118.7\pm4.1$ \\
    Maunder        & $3.8\pm3.7$    & $14\pm3$  & $8.78\pm1.75$    & $55.5\pm8.6$  & $1.66\pm0.15$  & $116.7\pm2.8$ \\
    \end{tabular}
    \caption{Linear-fit coefficients for the $P_{\mathrm{Rieg}}$--$P_{\mathrm{rot}}$ relation in each rotational regime. Uncertainties are from 300 bootstrap resamples.}
    \label{tab:regime_ab}
\end{table*}

The slow-rotator population additionally approaches a characteristic ratio

$$
X_{\mathrm{R}} = \frac{P_{\mathrm{Rieg}}}{P_{\mathrm{rot}}}\simeq 6.
$$

\begin{figure}[t]
    \centering
    \includegraphics[width=0.5\linewidth]{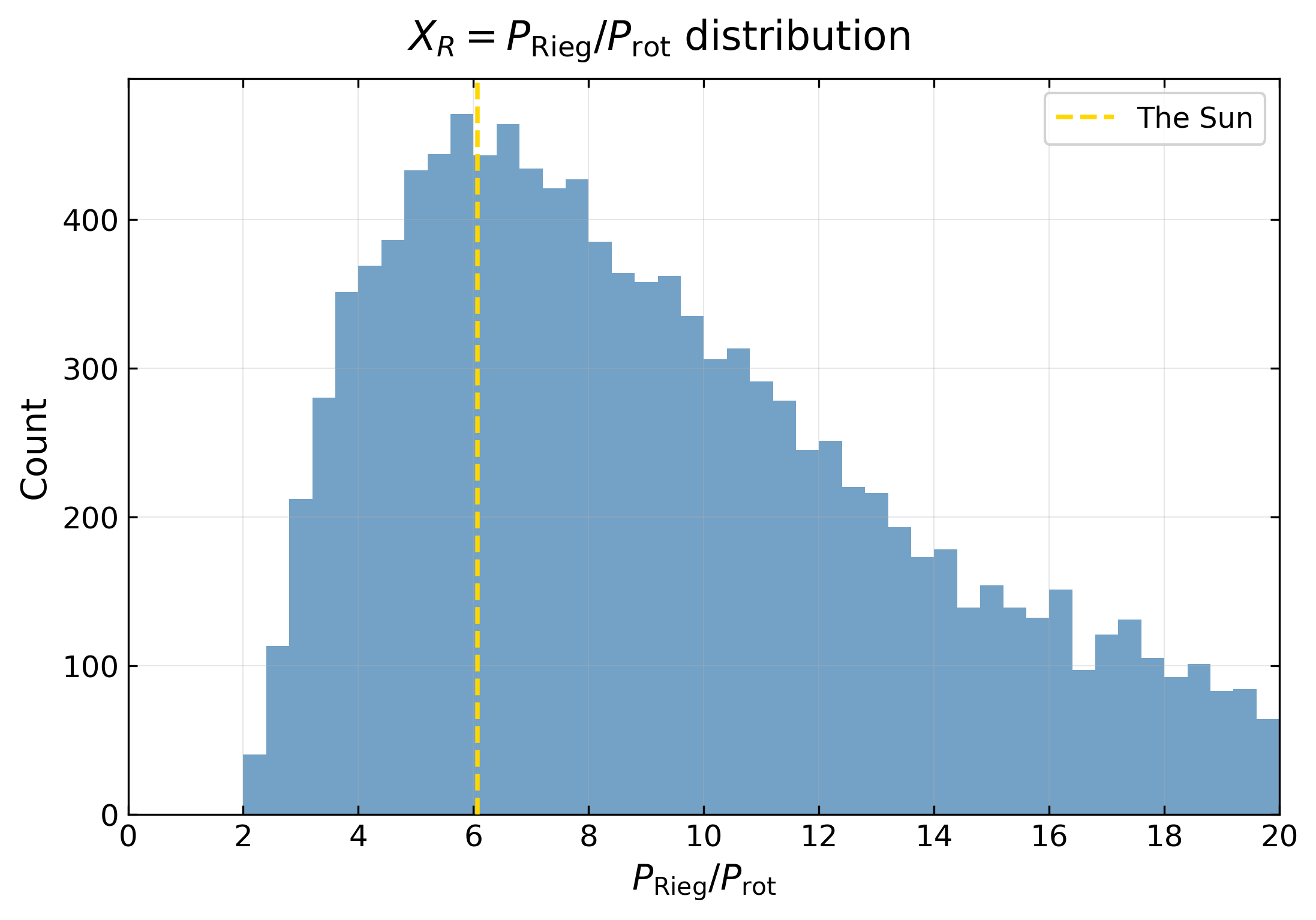}
    \caption{Distribution of the Maunder $X_{\mathrm{R}}$, excluding the anomalous population. Other catalogs demonstrate similar behavior.
}
    \label{fig:xr histogram}
\end{figure}

This value is close to the solar ratio, for which the canonical solar Rieger periodicity of approximately $154$~d corresponds to about six solar rotations. This correspondence is also evident in Figure~\ref{fig:xr histogram}, where the distribution of $X_{\mathrm{R}}$ shows a prominent peak near $X_{\mathrm{R}}\simeq6$. The dominance of this ratio in the overall distribution largely reflects the fact that most stars in our sample belong to the slow-rotator population, which is itself concentrated around $X_{\mathrm{R}}\simeq6$.

Thus, although the $P_{\mathrm{Rieg}}$--$P_{\mathrm{rot}}$ distribution contains three rotational regimes, these regimes exhibit only two distinct forms of period coupling. The intermediate rotational regime defines a strong period-coupling regime, in which $P_{\mathrm{Rieg}}$ increases steeply with $P_{\mathrm{rot}}$, whereas both the fast and slow rotational regimes belong to a weak period-coupling regime, in which $P_{\mathrm{Rieg}}$ depends only weakly on $P_{\mathrm{rot}}$. The three rotational regimes therefore map onto two distinct period-coupling regimes.

The stellar properties of the three rotational regimes are summarized in Table~\ref{tab:regimes_2}. The slow regime begins at $P_{\mathrm{rot}}\simeq7$--$9$~d and is dominated by relatively cool stars. Its median effective temperature is approximately $5270$~K, with the central $16$--$84\%$ interval extending from roughly $4400$ to $5900$~K. Only about $8\%$ of slow rotators have $T_{\mathrm{eff}}>6000$~K.

The intermediate regime occupies a systematically hotter part of the sample than the slow population, with a median effective temperature of approximately $5850$~K. Its temperature distribution is shifted toward higher values relative to the slow regime, indicating that the transition between the two rotational regimes is accompanied by a clear change in the underlying stellar population.

The fast regime is markedly different, with a median $T_{\mathrm{eff}}=7102$~K and a $16$--$84\%$ range of $6642$--$7502$~K; $97\%$ of these stars are hotter than $6000$~K. This temperature distribution places the population primarily in the F-type regime occupied by $\gamma$~Doradus stars. This identification is independently supported by cross-matching with the $\gamma$~Doradus catalogue of \citet{Li2020}: 50\% of the 407 fast-regime stars are present in their catalogue, compared with only 0.3\% of the remainder of our sample, and the fast regime contains 84\% of all \citet{Li2020} matches in our sample.

The stellar structure of these hot stars also differs substantially from that of the cooler late-type dwarfs considered below. They possess convective cores and relatively thin outer convective envelopes, so the conventional envelope-based convective turnover time used in our Rossby-number calibration cannot be straightforwardly applied. We therefore treat this population separately in the Rossby-number analysis (Section~\ref{sec:Ro}).

A further complication is that $\gamma$~Doradus stars exhibit high-order gravity-mode pulsations with characteristic periods of approximately $0.3$--$3$~d \citep{Balona2011,Li2020}, overlapping the timescales of rapid stellar rotation. The catalogued $P_{\mathrm{rot}}$ values may therefore trace surface rotation for some stars but pulsational variability for others. We retain these catalogued periods as the observational timescale defining the $P_{\mathrm{Rieg}}$--$P_{\mathrm{rot}}$ morphology, while treating their interpretation as surface rotation with caution for the hot fast-regime population.

The effective-temperature dependence therefore shows a clear progression across the three rotational regimes. Slow rotators are predominantly cool stars, the intermediate regime extends to systematically higher temperatures, and the fast regime is almost entirely composed of hot stars. In Section~\ref{sec:literature comparison}, we compare the observed $P_{\mathrm{Rieg}}$--$P_{\mathrm{rot}}$--$T_{\mathrm{eff}}$ behavior with that reported by \citet{Arkhypov2021}.

Although the rotational regimes show clear systematic differences in $T_{\mathrm{eff}}$, their temperature distributions overlap, and the two weakly coupled populations occupy opposite ends of the temperature distribution. Effective temperature alone therefore does not determine the period-coupling regime.

\begin{table*}[t]
    \centering
    \setlength{\tabcolsep}{3pt}
    \small
    \begin{tabular}{ll|c|c|cc|ccc|ccc}
    & & & & \multicolumn{2}{c|}{$P_{\mathrm{rot}}$ [d]} & \multicolumn{3}{c|}{$T_{\mathrm{eff}}$ [K]} & \multicolumn{3}{c}{$\mathrm{Ro}$} \\
    Catalog & Regime & Cut [d] & $N$ & median & range & median & 16--84\% & $>6000$ & median & 16--84\% & $>0.6$ \\
    \hline
    McQuillan 2014 & Intermediate & $<8.0$       &  1930 & $5.53$  & $0.32$--$8.00$ & $5849$ & $5314$--$6225$ & $37\%$ & $0.44$ & $0.29$--$0.83$ & $33\%$ \\
                   & Slow         & $\geq8.0$    &  9813 & $17.25$ & $8.01$--$65.8$ & $5274$ & $4415$--$5798$ & $8\%$  & $0.97$ & $0.59$--$1.36$ & $83\%$ \\
    \hline
    Reinhold 2023  & Intermediate & $<8.0$       &  1881 & $5.8$  & $1.08$--$7.97$ & $5827$ & $5284$--$6191$ & $35\%$ & $0.45$ & $0.29$--$0.84$ & $34\%$ \\
                   & Slow         & $\geq8.0$    & 10323 & $18.17$ & $8.07$--$46.7$ & $5274$ & $4404$--$5798$ & $8\%$  & $1.01$ & $0.61$--$1.37$ & $85\%$ \\
    \hline
    Santos 2021    & Intermediate & $<9.0$       &  2173 & $6.34$  & $0.99$--$8.98$ & $5887$ & $5470$--$6210$ & $39\%$ & $0.51$ & $0.34$--$0.92$ & $40\%$ \\
                   & Slow         & $\geq9.0$    &  5551 & $16.30$ & $9.03$--$72.4$ & $5568$ & $5286$--$5926$ & $12\%$ & $1.06$ & $0.69$--$1.43$ & $90\%$ \\
    \hline
    Maunder        & Fast         & $<1.5$       &   407 & $0.99$  & $0.42$--$1.51$ & $7102$ & $6642$--$7502$ & $97\%$ & $1.77$ & $0.85$--$3.23$ & $91\%$ \\
                   & Intermediate & $1.5$--$7.0$ &  1795 & $4.6$  & $1.51$--$7.00$ & $5985$ & $5300$--$6468$ & $49\%$ & $0.53$ & $0.28$--$1.35$ & $45\%$ \\
                   & Slow         & $\geq7.0$    & 10896 & $17.21$ & $7.00$--$48.0$ & $5258$ & $4418$--$5799$ & $8\%$  & $0.97$ & $0.58$--$1.33$ & $82\%$ \\
    \end{tabular}
    \caption{Rotation-period, effective-temperature, and Rossby-number properties of the rotational regimes. Rossby numbers for the fast regime are extrapolated beyond the calibrated range and should be interpreted with caution.}
    \label{tab:regimes_2}
\end{table*}

As an additional check on whether the observed Rieger-type behavior extends to stars with substantially different internal structures or magnetic properties, we examined several special stellar populations represented in our sample. We identify 116 fully convective M dwarfs, of which 96 lie in the slow regime. Their $P_{\mathrm{Rieg}}$ values span a broad range comparable to that of the other slow-regime stars. We also identify four Ap/Bp chemically peculiar stars \citep{Renson2009,Shang2022}; three of the four have $P_{\mathrm{Rieg}}<50$~d, and they are predominantly located in the intermediate regime.

\subsection{Fast-High-$P_{\mathrm{Rieg}}$: Tidally Synchronized Binaries}
\label{sec:Binaries}

The $P_{\mathrm{Rieg}}$--$P_{\mathrm{rot}}$ distribution reveals a distinct population of very rapid rotators with unusually long $P_{\mathrm{Rieg}}$, which was excluded from the rotational-regime analysis in Section~\ref{sec:RiegerRotationalRegimes}. We now examine this population separately and show that it is strongly associated with stellar multiplicity, and in particular with close, tidally synchronized binary systems.

Distinguishing genuinely young rapid rotators from older stars whose rapid rotation is maintained by tidal synchronization is a fundamental challenge in interpreting stellar rotation measurements \citep{KamaiPerets2025}. Previous approaches have relied on statistical age estimates from stellar kinematics and on the distribution of stars in a reduced-dimensional representation of their photometric properties \citep{Kamai2026}. Here, we investigate whether Rieger-type periodicities provide a physically motivated alternative for distinguishing these populations using photometry alone.

\begin{figure}[t]
\centering
\includegraphics[width=0.75\linewidth]{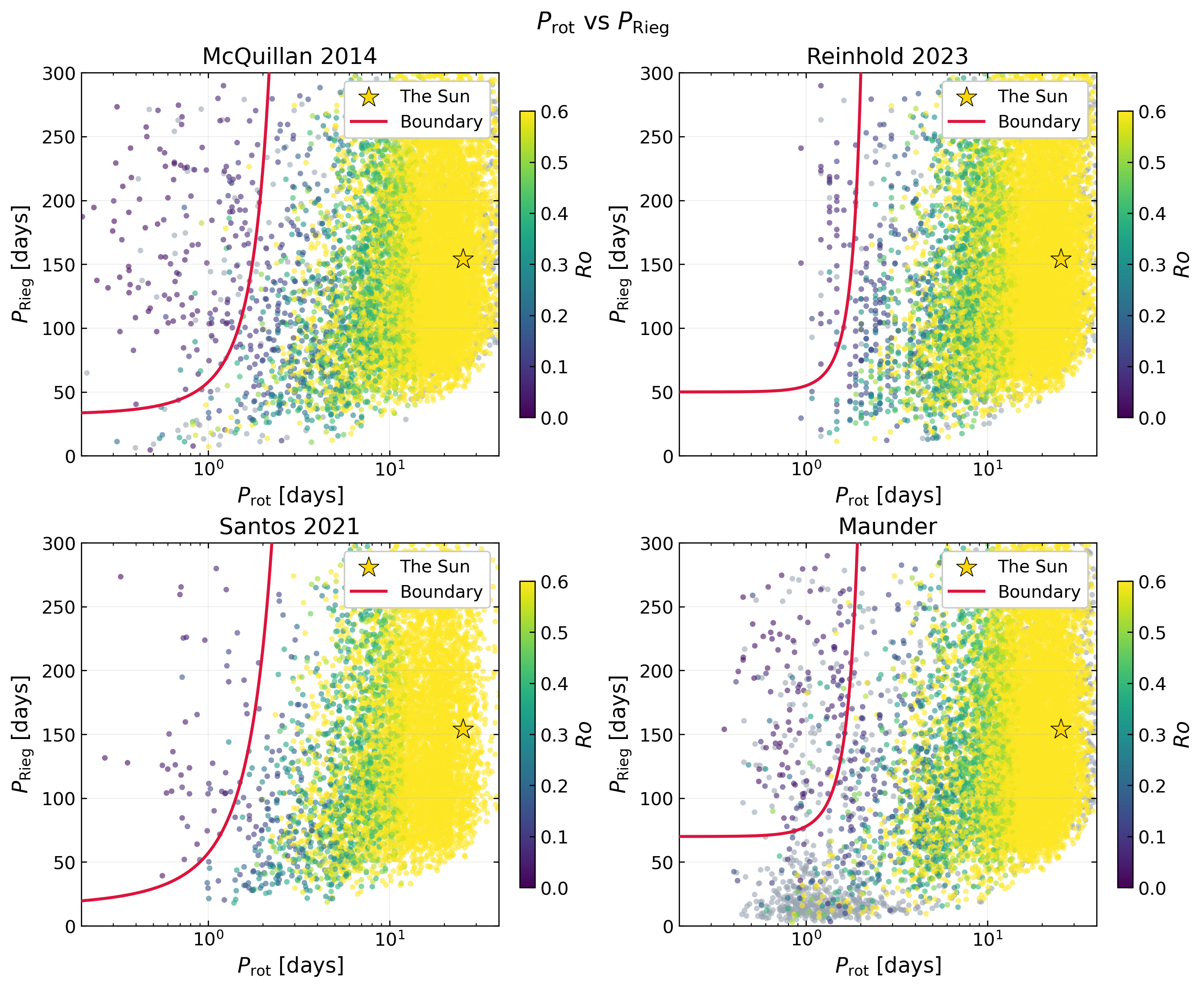}
\caption{$P_{\mathrm{Rieg}}$ versus $P_{\mathrm{rot}}$ on a logarithmic $P_{\mathrm{rot}}$ axis, colored by Rossby number. Red curves show the optimized binaries classification boundaries.}
\label{fig:Prieg clf}
\end{figure}

As shown in Figure~\ref{fig:Prieg clf}, the anomalous population is concentrated at $P_{\mathrm{rot}}\lesssim2.5$~d and has unusually low inferred Rossby numbers, with a median of $\mathrm{Ro}\simeq0.06$ and $79\%$ of the stars below $\mathrm{Ro}=0.15$. To isolate this low-Rossby, rapid-rotator population from the remaining stellar populations, including the hot rapid rotators, we searched for an optimized empirical boundary in the $P_{\mathrm{Rieg}}$--$P_{\mathrm{rot}}$ plane that separates the low-Rossby binary population from the single-star fast- and intermediate-rotator populations. The resulting boundary has the form

$$
P_{\mathrm{Rieg}} =
y_0 \exp\left[\left(\frac{P_{\mathrm{rot}}}{\lambda}\right)^n\right],
$$

and acts as a lower limit on $P_{\mathrm{Rieg}}$. It approaches $y_0$ for $P_{\mathrm{rot}}\ll\lambda$ and rises steeply at longer rotation periods. Because the rotation-period distributions differ between catalogs, the boundary is optimized separately for each catalog. The resulting boundary parameters, together with the binary fractions of the selected populations, are summarized in Table~\ref{tab:clump}.

\begin{table}[t]
    \centering
    \begin{tabular}{l|ccc|c|c|cc}
    Catalog & $y_0$ [d] & $\lambda$ [d] & $n$ & $P_{\mathrm{rot}}^{\mathrm{max}}$ [d] & $N$ & $f_{\mathrm{bin}}$ & $f_{\mathrm{bin, Ro}}$ \\
    \hline
    McQuillan 2014 & 32.5 & 1.375 & 1.75 & 2.1 & 217 & 61\% & 69\% \\
    Reinhold 2023  & 50 & 1.75 & 4.25 & 2.01 &  57 & 61\% & 73\% \\
    Santos 2021    & 15.0 & 0.750 & 1.00 & 2.25 &  60 & 65\% & 75\% \\
    Maunder        & 70 & 1.75 & 4 & 1.92 & 271 & 66\% & 80\% \\
    \hline
    Pooled & 60 & 1.875 & 4.5 & 2.08 & 654 & 63\% & 72\% \\
    \end{tabular}
    \caption{Optimal parameters of the empirical boundary used to isolate the anomalous low-Rossby rapid-rotator population, identified predominantly with binaries, from the remaining stellar populations. The low-Rossby fraction is the fraction of selected stars with a measured Rossby number that satisfy $\mathrm{Ro}<0.1$. Stars without a defined Rossby number are retained in the fitting procedure and treated as non-binaries. The quantities $f_{\mathrm{bin}}$ and $f_{\mathrm{bin,Ro}}$ denote the fractions of selected stars flagged as binaries by \citet{GodoyRivera2025} in the full selected sample and among stars with a defined Rossby number, respectively.}
    \label{tab:clump}
\end{table}

Cross-matching the populations selected by these boundaries with the binary classifications of \citet{GodoyRivera2025} shows a strong enhancement in multiplicity (Table~\ref{tab:clump}). Across the individual rotation-period catalogs, $61$--$66\%$ of all selected stars are flagged as binaries, with a pooled fraction of $f_{\mathrm{bin}}=63\%$, compared with only $22\%$ of the full stellar sample. The association becomes still stronger when considering only selected stars for which a Rossby number is available: $f_{\mathrm{bin,Ro}}$ ranges from $69\%$ to $80\%$ across the catalogs and is $72\%$ in the pooled sample. Thus, the population isolated by the $P_{\mathrm{Rieg}}$--$P_{\mathrm{rot}}$ boundary is not simply a sample of rapid rotators, but is strongly enriched in binaries, particularly within the subset for which its anomalously low Rossby numbers can be established directly. Importantly, this enhancement is not reproduced by either coordinate alone: fast rotators outside the boundary have substantially lower binary fractions, while dividing the sample using $P_{\mathrm{Rieg}}$ alone produces only a weak contrast. The combination of $P_{\mathrm{rot}}$ and $P_{\mathrm{Rieg}}$ therefore provides considerably stronger discriminatory power between these anomalous systems and the genuine rapid-rotator population.

Having identified the anomalous population as strongly associated with binaries, we next ask whether these systems are specifically consistent with tidal synchronization. Rapidly rotating stars in close binaries are natural candidates for synchronization, whereas wide binaries should not show the same behavior. The orbital information in the \citet{GodoyRivera2025} catalog provides a direct way to distinguish between these possibilities. Among stars carrying close or interacting binary flags, $18.2\%$ lie inside the boundary, compared with only $3.3\%$ of systems carrying exclusively wide or astrometric flags, essentially identical to the $3.1\%$ occurrence rate in the full sample. Similarly, $51.9\%$ of eclipsing binaries with $P_{\mathrm{orb}}<3$~d lie inside the boundary, compared with only $1.0\%$ for $P_{\mathrm{orb}}>10$~d. For boundary stars with independently measured orbital periods, the measured photometric rotation also closely follows the orbital period, with a median $P_{\mathrm{rot}}/P_{\mathrm{orb}}=0.94$. The anomalous population is therefore associated specifically with close systems in which the observed rotation has been strongly affected by tidal synchronization.

This concentration of tidally synchronized binaries at very low Rossby number is consistent with \citet{Mathur2025}, who noted that the parameter space corresponding to the saturated regime, at $\mathrm{Ro} < 0.13$, in the \emph{Kepler} sample is populated by binaries, particularly tidally synchronized systems.

As summarized in Table~\ref{tab:tidalprops}, the stellar properties of these systems provide an additional and independent clue to their origin. Despite their observed rotation periods of approximately one day, the binary population closely resembles the ordinary slow rotators in effective temperature, mass, metallicity, surface gravity, radius, and, most importantly, $P_{\mathrm{Rieg}}$. The median $P_{\mathrm{Rieg}}$ of the synchronized-binary population is $162.6$~d, compared with $149.7$~d for the ordinary slow-rotator population, whereas genuine single fast rotators have a median $P_{\mathrm{Rieg}}$ of only $18$~d and are substantially hotter and more massive. The intermediate-regime stars generally lie between the slow and genuine fast populations in these quantities. The synchronized binaries therefore do not continue the normal slow--intermediate--fast progression; instead, they retain the properties of the slow population while appearing at extremely short measured $P_{\mathrm{rot}}$. 

We do not identify a distinct analogous binary population whose stellar properties resemble those of the intermediate regime. Such a population may be obscured by the intrinsic scatter and measurement uncertainties, may be comparatively rare, or may not form a sufficiently distinct structure in the present data.

\begin{table}[t]
    \centering
    \begin{tabular}{l|cccc}
    Property & Binaries & Slow & Intermed. & Single fast \\
     & ($367$) & ($11\,219$) & ($3\,294$) & ($439$) \\
    \hline
    $T_{\mathrm{eff}}$ [K]  & 5472    & 5242    & 5818    & 7063    \\
    $M$ [$M_\odot$]         & 0.802   & 0.822   & 0.949   & 1.466   \\
    $[\mathrm{Fe/H}]$       & $-0.20$ & $-0.10$ & $-0.16$ & $-0.14$ \\
    $P_{\mathrm{Rieg}}$ [d] & 162.6   & 149.7   & 106.6   & 18.0    \\
    $\log g$                & 4.54    & 4.55    & 4.45    & 4.15    \\
    $R$ [$R_\odot$]         & 0.78    & 0.81    & 0.98    & 1.66    \\
    $\mathrm{Ro}$           & 0.063   & 1.016   & 0.49    & 1.63    \\
    $P_{\mathrm{rot}}$  [d] & 0.94    & 18.3    & 5.99    & 1.0     \\
    \end{tabular}
    \caption{Median properties of the four rotational populations of the pooled catalog, with the number of stars in each given below the column head. Rossby numbers for the single fast-rotator population are extrapolated beyond the calibrated range and should therefore be interpreted with caution.}
    \label{tab:tidalprops}
\end{table}

The Rossby number is the main exception to the otherwise close agreement between the synchronized binaries and the slow rotators. This is expected because $\mathrm{Ro}$ is calculated directly from the observed $P_{\mathrm{rot}}$. To estimate the rotation periods these systems would have in the absence of tidal synchronization, we matched the stellar properties of the synchronized-binary population to those of the full non-binary population and determined the corresponding location along the $P_{\mathrm{rot}}$ axis. This yields a characteristic rotation period of approximately $18$~d, compared with the observed median of $0.94$~d, with a median star-by-star ratio of approximately $14$. We perform this correction only statistically at the population level, rather than for individual stars, because stellar properties such as $T_{\mathrm{eff}}$ show substantial spread at fixed $P_{\mathrm{rot}}$. Applying this factor shifts the median Rossby number of the synchronized binaries from $0.06$ to $0.93$, essentially identical to the median $\mathrm{Ro}=1.016$ of the slow rotators.

\subsection{Comparison with Arkhypov et al.}
\label{sec:literature comparison}

Having characterized the four populations in the $P_{\mathrm{Rieg}}$--$P_{\mathrm{rot}}$ plane, we next compare their population-level scaling with the previous large-sample stellar analysis of \citet{Arkhypov2021}. Our final sample contains 15,319 unique stars with measured Rieger-type periods, nearly nine times the size of their 1,726-star \emph{Kepler} sample. Their analysis differed substantially from ours: it used the $A_1^2$ variability metric, rotation periods from the McQuillan catalog, and a period search spanning $2P_{\mathrm{rot}}$ to 1000~d. Importantly, \citet{Arkhypov2021} did not assign an individual Rieger-type period to each star. Instead, they measured a dominant activity period, $P_{\mathrm{peak}}$, for each target and identified the Rieger-type relation from the lower envelope of the $P_{\mathrm{peak}}$--$P_{\mathrm{rot}}$ distribution, motivated by the location of the solar Rieger periodicity near the lower boundary of their stellar distribution. Despite these methodological differences, their study provides the most direct population-level comparison with the relation measured here.

\citet{Arkhypov2021} defined this lower-envelope timescale, $P_{\mathrm{RTC}}$, by taking the minimum $P_{\mathrm{peak}}$ in bins of $P_{\mathrm{rot}}$. They then quantified its temperature dependence through the fitted slope $D$ of the $\log P_{\mathrm{RTC}}$--$\log P_{\mathrm{rot}}$ relation. They reported three approximate regimes: $D\simeq0$ for cool stars with $3000\lesssim T_{\mathrm{eff}}\lesssim4000$~K, $D\simeq0.35$ at intermediate temperatures, and $D$ approaching unity for the hottest interval, approximately $6300\lesssim T_{\mathrm{eff}}\lesssim7000$~K.

A direct comparison of the lower envelopes is not appropriate because the solar Rieger periodicity lies near the center of our $P_{\mathrm{Rieg}}$ distribution rather than its lower boundary. As shown in Section~\ref{sec:rieger_bias_results}, this difference is largely associated with the different treatment of long-timescale variability: removing our detrending step shifts the recovered distribution toward longer periods and places the solar value much closer to its lower envelope. We therefore compare the temperature-dependent scaling rather than the absolute lower-envelope location.

We therefore compare the temperature dependence of the two analyses by fitting a power-law relation,

$$
\log(P_{\mathrm{Rieg}}) = D\log(P_{\mathrm{rot}}) + C,
$$

within different effective-temperature intervals. For this comparison, we exclude the anomalous binary population using the fitted boundary. We restrict the comparison to $0.5\le P_{\mathrm{rot}}\le30$~d, matching the rotation-period range adopted by \citet{Arkhypov2021}. The fitted slopes are listed in Table~\ref{tab:rieger_D}. We show only the McQuillan and Maunder rotation catalogs, since the Reinhold and Santos catalogs contain too few stars in the coolest and hottest temperature intervals to constrain the corresponding slopes reliably.

\begin{table}[t]
    \centering
    \begin{tabular}{l|cc|cc|c}
    & \multicolumn{2}{c|}{McQuillan 2014} & \multicolumn{2}{c|}{Maunder} & AK21 \\
    $T_{\mathrm{eff}}$ [K] & $D$ & $N$ & $D$ & $N$ & $D$ \\
    \hline
    3000--4000 & $+0.07\pm0.06$ &  534 & $+0.17\pm0.05$ &  614 & $-0.07\pm0.07$ \\
    4000--5000 & $+0.11\pm0.02$ & 2416 & $+0.22\pm0.02$ & 2811 & \dots           \\
    5000--5500 & $+0.19\pm0.02$ & 3251 & $+0.2\pm0.02$ & 3572 & \dots           \\
    5500--6000 & $+0.28\pm0.02$ & 3381 & $+0.29\pm0.02$ & 3352 & $+0.34\pm0.09$ \\
    6000--6300 & $+0.32\pm0.03$ &  1213 & $+0.31\pm0.03$ &  1248 & \dots           \\
    6300--7000 & $+0.71\pm0.05$ &  263 & $+0.8\pm0.03$ &  528 & $+0.85\pm0.09$ \\
    \end{tabular}
    \caption{Power-law index $D$ of the $P_{\mathrm{Rieg}}$--$P_{\mathrm{rot}}$ relation in different effective-temperature intervals. $N$ denotes the number of stars in each fit, and AK21 gives the corresponding values reported by \citet{Arkhypov2021}. The fit is calculated for stars after removing the binary populations using the fitted boundary.}
    \label{tab:rieger_D}
\end{table}

Despite the substantial differences between the two analysis methods, the resulting temperature-dependent trends are in good agreement with those reported by \citet{Arkhypov2021} (Table~\ref{tab:rieger_D}). Both analyses recover a weak or nearly rotation-independent Rieger-type scaling among the coolest stars, a progressively stronger dependence at intermediate temperatures, and a steep, nearly rotation-proportional relation in the hottest interval. The agreement across the McQuillan and Maunder samples indicates that this systematic increase in the $P_{\mathrm{Rieg}}$--$P_{\mathrm{rot}}$ dependence with effective temperature is robust to the different approaches used to identify the Rieger-type timescale.

For this comparison, we adopt $6300$--$7000$~K as the hottest temperature interval and do not extend the analysis to higher temperatures. This both limits the contribution of the very hot fast-rotator population uniquely resolved by the Maunder catalog and preserves a like-for-like comparison with \citet{Arkhypov2021}. AK21 used rotation periods from the McQuillan catalog and likewise stopped at $7000$~K; consequently, extending our comparison to hotter stars would introduce a population that was not represented in their analysis.

\subsection{Intermediate–Slow Rossby-Number Transition}
\label{sec: Ro Transition}

The comparison above confirms that the strength of the $P_{\mathrm{Rieg}}$--$P_{\mathrm{rot}}$ relation changes systematically across the stellar population, but it does not establish which physical parameter organizes this change. We therefore next examine the same transition, which appears in the $P_{\mathrm{Rieg}}$--$P_{\mathrm{rot}}$ plane as the boundary between the intermediate and slow regimes, as a function of Rossby number, restricting the analysis to main-sequence stars within the calibrated range of the adopted \citet{Noyes1984a} relation. The hot fast-rotator population discussed above is therefore excluded from the quantitative Rossby-number analysis.

Within the calibrated sample, the slow rotational regime lies predominantly at higher Rossby number (Table~\ref{tab:regimes_2}). Depending on the rotation-period catalog, its median Rossby number is $\mathrm{Ro}\simeq0.97$--$1.06$, and $82$--$90\%$ of slow rotators with available Rossby numbers have $\mathrm{Ro}>0.6$. In contrast, the intermediate rotational regime is concentrated toward lower Rossby numbers, with median $\mathrm{Ro}\simeq0.44$--$0.53$. This change in Rossby number coincides with the change in the $P_{\mathrm{Rieg}}$--$P_{\mathrm{rot}}$ relation from the strongly rotation-dependent intermediate regime to the much more weakly rotation-dependent slow regime.

The Rossby-number distributions of the intermediate and slow populations are not completely disjoint, and the transition is therefore statistical rather than a sharp division between individual stars. Nevertheless, the change from strong to weak $P_{\mathrm{Rieg}}$--$P_{\mathrm{rot}}$ coupling is clearly associated with an increase in Rossby number within the calibrated late-type stellar sample. Whether Rossby number provides a more fundamental description of this transition than the correlated quantities $P_{\mathrm{rot}}$, $T_{\mathrm{eff}}$, and metallicity is examined separately below.

To quantify the Rossby-number transition, we restricted the analysis to stars with $P_{\mathrm{rot}}>3.5$~d, thereby isolating the intermediate-to-slow transition. We then divided the sample into Rossby-number intervals and, within each interval, fitted the $P_{\mathrm{Rieg}}$--$P_{\mathrm{rot}}$ relation to obtain the corresponding power-law slope $\mathrm{a}$. Each fitted slope was assigned to the central Rossby number of its interval, producing $\mathrm{a}$ as a function of $\mathrm{Ro}$. We then fitted this relation with a broken-line function and identified the break point as the characteristic transition Rossby number. This procedure was applied independently to each rotation-period catalog and to the pooled sample. The resulting catalog-specific transition values are listed in Table~\ref{tab:ro_threshold_per_catalog}.

\begin{figure}[t]
    \centering
    \includegraphics[width=0.5\linewidth]{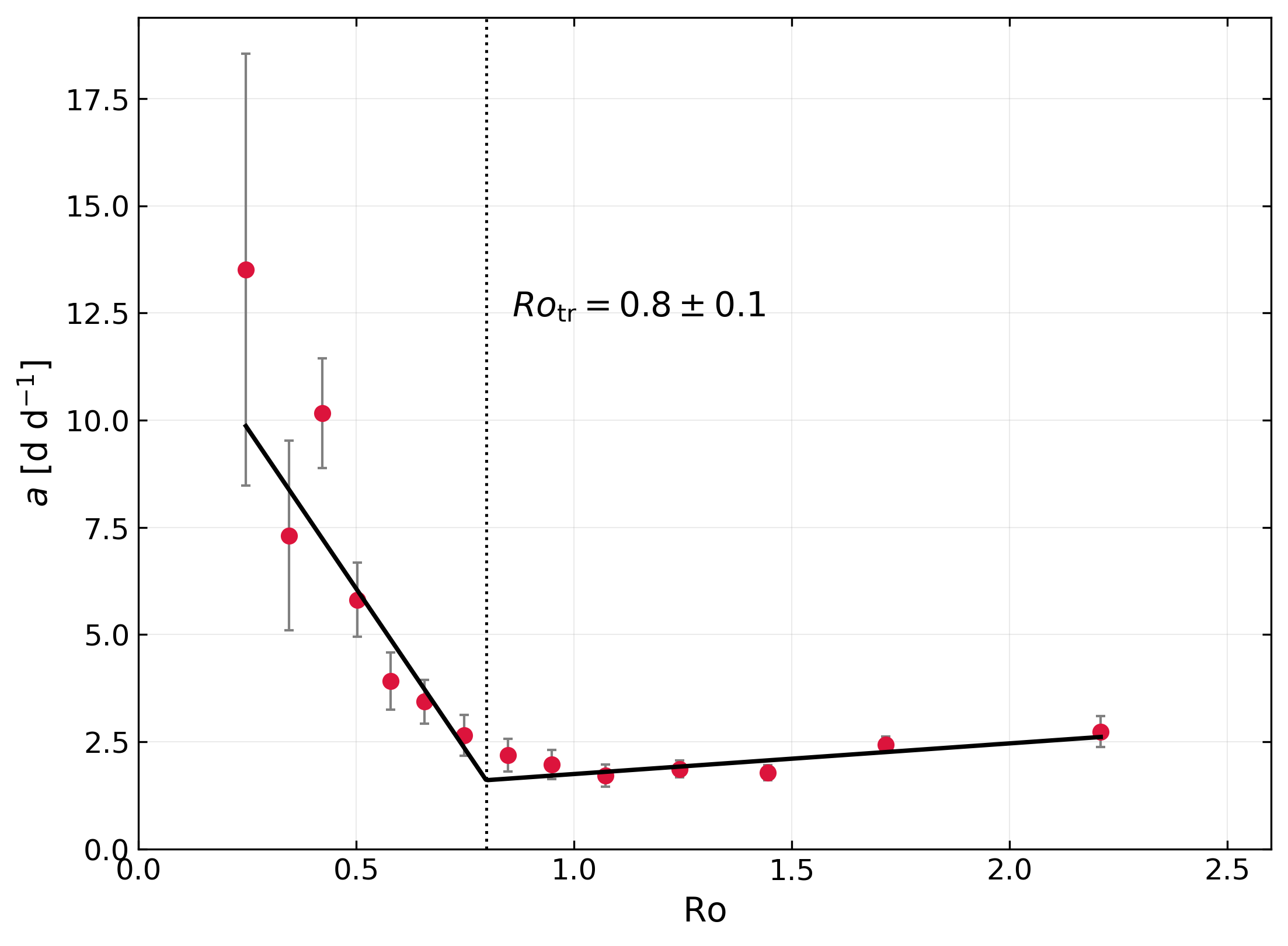}
    \caption{
    Local $P_{\mathrm{Rieg}}$--$P_{\mathrm{rot}}$ slope as a function of Rossby number for the pooled sample. The solid line shows the two-segment fit and the dotted line marks $\mathrm{Ro}_{\mathrm{tr}}$.
}
    \label{fig:ro fit}
\end{figure}

The catalog-specific estimates span $\mathrm{Ro}_{\mathrm{tr}}\simeq0.70$--$0.90$, indicating some catalog-to-catalog variation in the inferred transition. Nevertheless, the uncertainty ranges of the individual estimates overlap that of the pooled result of $\mathrm{Ro}_{\mathrm{tr}}\simeq0.8$. Given this overall consistency and the larger combined sample, we adopt the pooled value as our representative empirical transition, while noting the remaining catalog-dependent scatter. As shown in Figure~\ref{fig:ro fit}, this gives

$$
\mathrm{Ro}_{\mathrm{tr}} = 0.8 \pm 0.1.
$$

This transition provides an empirical boundary between the strongly coupled intermediate regime and the weakly coupled slow regime.

\begin{table}[t]
    \centering
    \begin{tabular}{l|c}
    Catalog & $\mathrm{Ro}_{\mathrm{tr}}$ \\
    \hline
    Pooled                     & $0.8\pm0.1$ \\
    \hline
    McQuillan 2014             & $0.7\pm0.12$ \\
    Reinhold 2023              & $0.7\pm0.1$ \\
    Santos 2021                & $0.9\pm0.08$ \\
    Maunder                    & $0.9\pm0.13$ \\
    \end{tabular}
    \caption{Rossby-number transition $\mathrm{Ro}_{\mathrm{tr}}$ for the pooled sample and individual rotation-period catalogs.}
    \label{tab:ro_threshold_per_catalog}
\end{table}

Importantly, the Rossby transition does not simply separate two regimes characterized by different but approximately constant values of the coupling slope $\mathrm{a}$. Instead, the behavior of $\mathrm{a}$ with Rossby number differs on the two sides of the transition. In the strongly coupled, low-Rossby-number regime, $\mathrm{a}$ increases systematically toward smaller $\mathrm{Ro}$, indicating that the dependence of $P_{\mathrm{Rieg}}$ on $P_{\mathrm{rot}}$ becomes progressively stronger as Rossby number decreases. In contrast, in the weakly coupled, high-Rossby-number regime, $\mathrm{a}$ shows little systematic variation with $\mathrm{Ro}$ and remains comparatively small. The transition therefore marks not only a change in the strength of the $P_{\mathrm{Rieg}}$--$P_{\mathrm{rot}}$ coupling, but also a change in how that coupling itself responds to Rossby number.

We additionally tested the sensitivity of the inferred transition to the adopted convective-turnover-time calibration. For the Santos catalog, our Noyes-based Rossby numbers give $\mathrm{Ro}_{\mathrm{tr}}=0.90\pm0.08$. Repeating the same transition analysis using the YREC-based Rossby numbers of \citet{Mathur2025}, which use the same Santos rotation periods but an independently derived convective turnover time, gives $\mathrm{Ro}_{\mathrm{tr}}=0.70\pm0.07$. Thus, changing the Rossby-number calibration shifts the precise numerical location of the transition only modestly, while preserving the transition within the same general $\mathrm{Ro}\simeq0.7$--$0.9$ range. Together with the consistency across the independent rotation-period catalogs (Table~\ref{tab:ro_threshold_per_catalog}), this shows that the inferred transition is not specific to either a single rotation catalog or a single convective-turnover-time prescription.

To assess whether Rossby number is simply correlated with the observed transition or instead provides the more natural coordinate for it, we first compare the transition with several closely related stellar parameters. The break in the $P_{\mathrm{Rieg}}$--$P_{\mathrm{rot}}$ relation appears near $\mathrm{Ro}\simeq0.8$, $P_{\mathrm{rot}}\simeq9$~d, and $T_{\mathrm{eff}}\simeq5400$~K, and these quantities are not independent, since $\mathrm{Ro}=P_{\mathrm{rot}}/\tau_c$ while the adopted convective turnover time is itself derived from stellar color and is therefore closely related to $T_{\mathrm{eff}}$. A transition associated primarily with one of these quantities can consequently leave a corresponding signature in the others, so the location of the break alone is not sufficient to identify the dominant control parameter. Instead, we test which parameter most clearly separates the two distinct $P_{\mathrm{Rieg}}$--$P_{\mathrm{rot}}$ coupling behaviors.

The more discriminating test is therefore whether the coupling itself changes across a candidate transition. We compare the fitted slope $a$ of the $P_{\mathrm{Rieg}}$--$P_{\mathrm{rot}}$ relation on the two sides of transitions defined in each stellar parameter. For $T_{\mathrm{eff}}$, $\tau_c$, mass, radius, metallicity, $\log g$, and $B-V$, the fitted values of $a$ before and after the putative transition remain very similar, such that little or no change in the $P_{\mathrm{Rieg}}$--$P_{\mathrm{rot}}$ coupling is visible when the sample is organized by these quantities. In contrast, dividing the sample by either $P_{\mathrm{rot}}$ or Rossby number produces a pronounced change in $a$, corresponding to the transition between the strongly and weakly rotation-coupled regimes. Thus, although several stellar parameters are correlated with the location of the transition, they do not independently reproduce the change in coupling that defines it. The transition is therefore primarily associated with either rotation period or Rossby number, rather than with $T_{\mathrm{eff}}$, surface gravity, metallicity, or the other structural parameters considered.

The main remaining ambiguity is therefore between $P_{\mathrm{rot}}$ and Rossby number, since the two are intrinsically correlated. To distinguish between rotation period and Rossby number, we first examine whether the Rieger-type timescale retains a dependence on stellar convection when rotation is approximately fixed. If $P_{\mathrm{rot}}$ alone controlled the transition, stars with the same rotation period should show no systematic dependence on the convective turnover time $\tau_c$; instead, we find a highly significant dependence of $P_{\mathrm{Rieg}}$ on $\tau_c$ at fixed $P_{\mathrm{rot}}$, detected at $10.6\sigma$, demonstrating that rotation alone is insufficient to describe the observed behavior.

We then generalize this comparison by defining a family of transition coordinates, $X_k=P_{\mathrm{rot}}/\tau_c^k$, for which $k=0$ corresponds to pure rotation-period control and $k=1$ corresponds to the standard Rossby number. Allowing the transition to select the value of $k$ that best separates the two $P_{\mathrm{Rieg}}$--$P_{\mathrm{rot}}$ branches gives $k=0.60\pm0.34$ from the star-level bootstrap, with similar preferred values obtained independently in each rotation-period catalogue; the preferred coordinate therefore includes a clear contribution from the convective turnover time and provides a better separation of the branches than $P_{\mathrm{rot}}$ alone, while remaining consistent with the Rossby scaling, $k=1$, within $1.2\sigma$. The improvement over pure rotation is modest, however, and the broad uncertainty on $k$ prevents the present data from uniquely identifying the standard Rossby scaling from this test alone.

Although the hot fast-rotator ($\gamma$~Doradus) population was excluded from the calibrated Rossby-number analysis above, its period-coupling behavior provides a useful qualitative extension of the result. Despite their substantially shorter cataloged $P_{\mathrm{rot}}$, these stars exhibit similarly weak $P_{\mathrm{Rieg}}$--$P_{\mathrm{rot}}$ coupling to that of the slow high-Rossby-number population, in contrast to the strong coupling observed in the intermediate regime, further supporting the conclusion that $P_{\mathrm{rot}}$ alone is insufficient to characterize the observed coupling regimes. Formally extrapolating the adopted Rossby-number calibration yields a median $\mathrm{Ro}=1.77$, with a $16$--$84\%$ range of $0.85$--$3.23$ (Table~\ref{tab:regimes_2}), placing most of the extrapolated values above the measured transition at $\mathrm{Ro}_{\mathrm{tr}}\simeq0.8$. This extrapolation should not be interpreted as a calibrated Rossby-number measurement for $\gamma$~Doradus stars, but it is qualitatively consistent with their weak $P_{\mathrm{Rieg}}$--$P_{\mathrm{rot}}$ coupling corresponding to the high-Rossby-number side of the transition.

Taken together, the dependence of $P_{\mathrm{Rieg}}$ on $\tau_{\mathrm{c}}$ at fixed $P_{\mathrm{rot}}$, the preference for a combined $P_{\mathrm{rot}}$--$\tau_{\mathrm{c}}$ coordinate, and the similarly weak rotational coupling of the hot fast-rotator and slow-rotator populations despite their substantially different rotation periods all support the interpretation that Rossby number provides a more natural coordinate for the observed transition than rotation period alone.

We caution that the inferred Rossby-number transition depends on several measured and model-dependent quantities, including $P_{\mathrm{rot}}$, $P_{\mathrm{Rieg}}$, and the adopted convective-turnover-time calibration. Measurement uncertainties, catalog-dependent systematics, and selection effects can therefore propagate into the inferred transition. Nevertheless, the transition remains within the same general $\mathrm{Ro}\simeq0.7$--$0.9$ range across the independent rotation-period catalogs and when using the alternative YREC-based Rossby numbers of \citet{Mathur2025}. In addition, the selection effects associated with the $P_{\mathrm{Rieg}}$ measurements are assessed and mitigated as described in Section~\ref{sec:rieger_bias_results}.

\subsection{Connection Between $P_{\mathrm{Rieg}}$ and Spot-Pattern Lifetime}
\label{sec:spot_lifetime}

Having established how $P_{\mathrm{Rieg}}$ is organized across the stellar populations, we next ask whether this intermediate magnetic timescale is reflected in the persistence of surface magnetic structures. We compare the measured Rieger-type periods with an independent estimate of the characteristic starspot-pattern lifetime derived from the light-curve autocorrelation function (ACF) using the calibration of \citet{Basri2022}. Although commonly referred to as a spot lifetime, this quantity more precisely represents the coherence time of the rotationally modulated surface pattern rather than the lifetime of an individual spot.

\begin{figure}[t]
    \centering
    \includegraphics[width=0.75\linewidth]{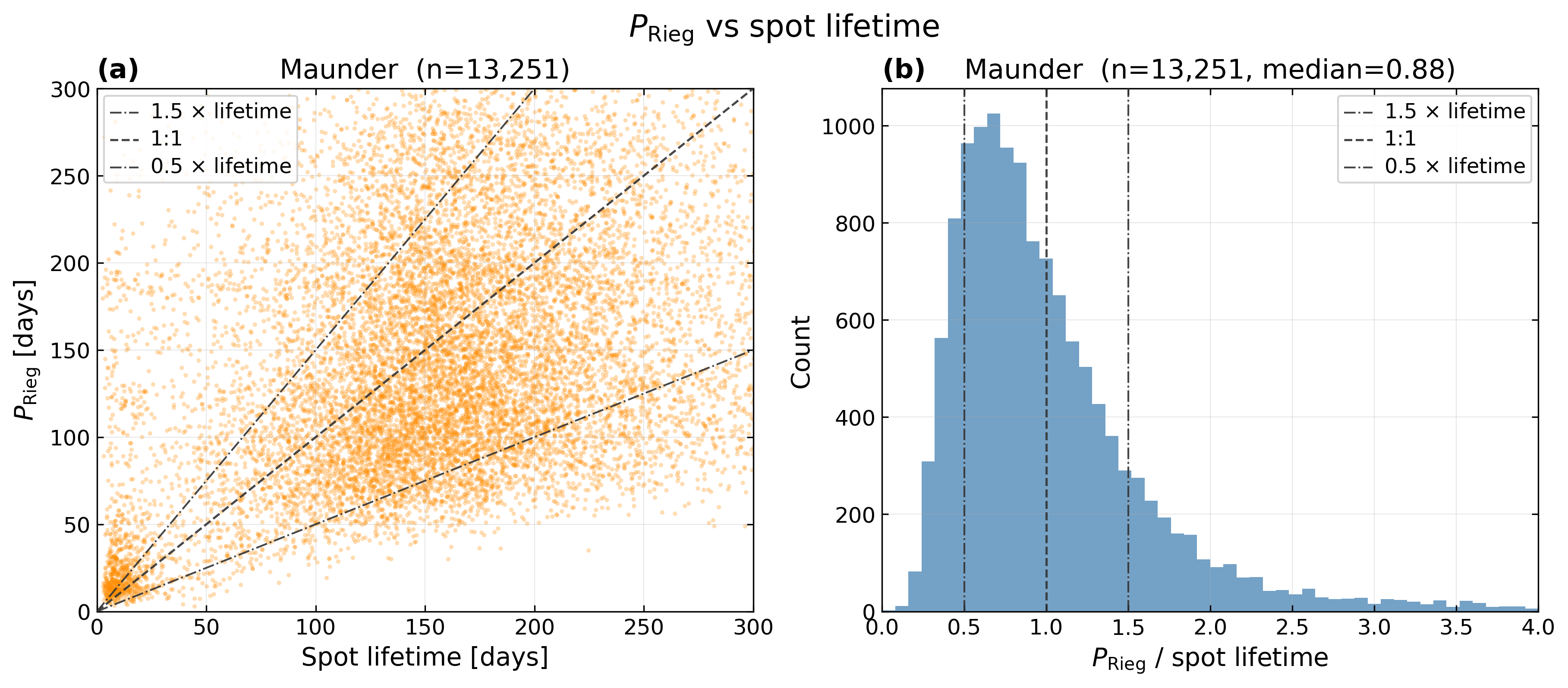}
    \caption{Comparison between $P_{\mathrm{Rieg}}$ and the spot lifetime. Panel (a) shows the $P_{\mathrm{Rieg}}$--spot-lifetime relation. Panel (b) shows the distribution of the ratio $P_{\mathrm{Rieg}}/L_{\mathrm{spot}}$.}
    \label{fig:life-time}
\end{figure}

We find that this coherence timescale is closely related to $P_{\mathrm{Rieg}}$, as shown in Figure~\ref{fig:life-time}. The ratio $P_{\mathrm{Rieg}}/L_{\mathrm{spot}}$ is concentrated near unity, with a median value of $0.88$ and a mode of $0.68$. The 16th--84th percentile interval is $0.51$--$1.58$, indicating that for most stars the two timescales agree to within approximately a factor of two.

\subsection{Selection Effects and Robustness}
\label{sec:rieger_bias_results}

\begin{figure}[t]
    \centering
    \includegraphics[width=0.75\linewidth]{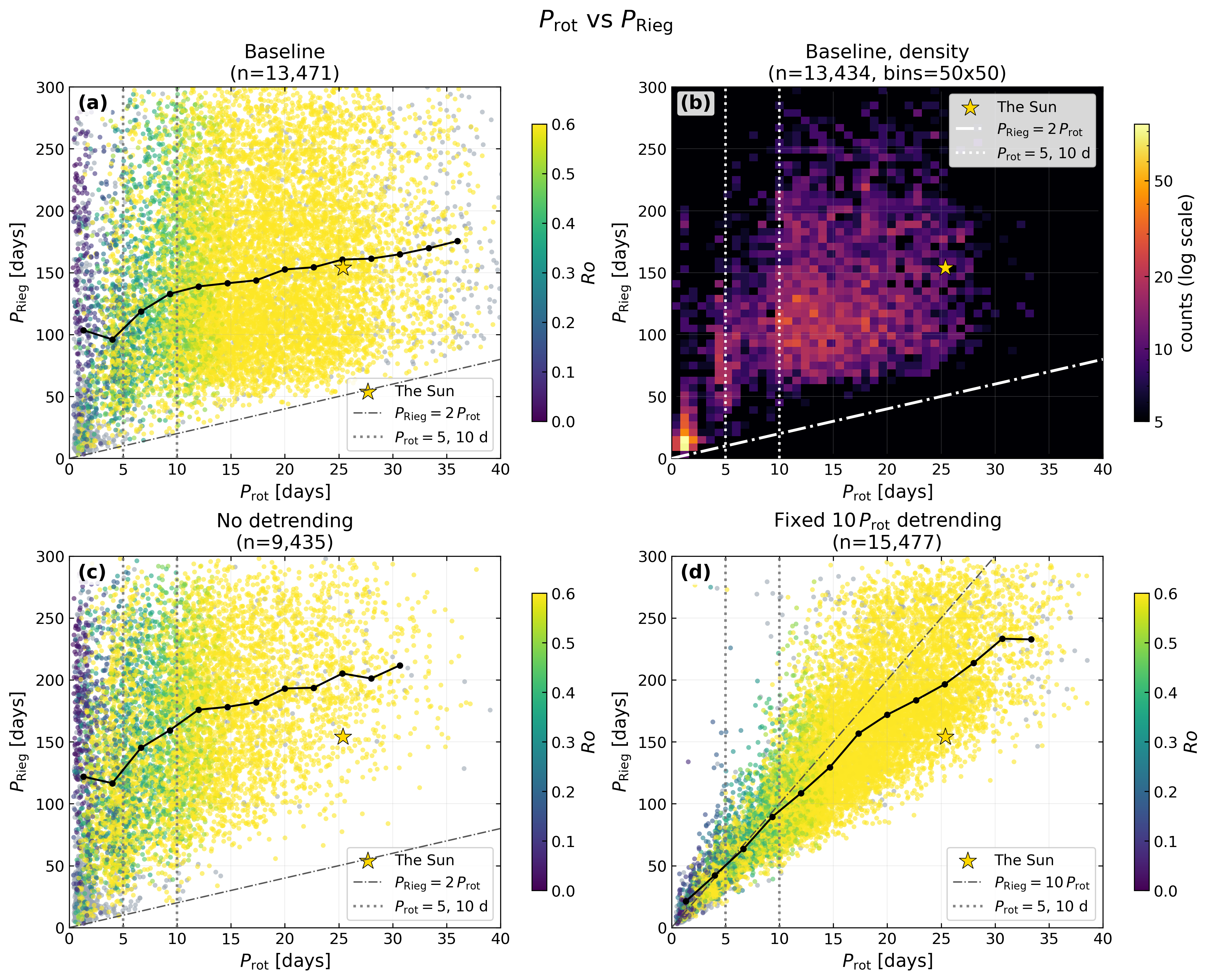}
    \caption{Comparison of the full pipeline (a, b), no detrending (c), and fixed $10P_{\mathrm{rot}}$ detrending (d) for the Maunder catalog. Panels (a), (c), and (d) show scatter plots, while panel (b) presents a density plot of the full pipeline results. Black curves indicate the binned medians, excluding the anomalous population. Reference lines mark $P_{\mathrm{Rieg}} = 2P_{\mathrm{rot}}$ and $P_{\mathrm{rot}} = 5$ and $10$ days. The density color scale in panel (b) starts at 5 counts per bin to exclude low-density bins.
}
    \label{fig:Prieg vs Prot color Ro no filter}
\end{figure}

Because the population morphology, regime transitions, and characteristic $P_{\mathrm{Rieg}}$ values described above could in principle be affected by the period-search boundaries or detrending procedure, we finally test the principal measurement and selection effects of the pipeline. First, we examined the influence of the detrending filter on the observed $P_{\mathrm{Rieg}}$--$P_{\mathrm{rot}}$ relation. For comparison, we applied an alternative version of the pipeline without detrending, as shown in Figure~\ref{fig:Prieg vs Prot color Ro no filter}. The resulting distribution exhibits a similar overall morphology and the same qualitative dependence of $P_{\mathrm{Rieg}}$ on $P_{\mathrm{rot}}$, but is systematically shifted toward longer periods. Without detrending, the solar Rieger periodicity lies close to the lower boundary of the stellar distribution, similar to the behavior reported by \citet{Arkhypov2021}. By contrast, the detrending procedure suppresses long-timescale variability that can dominate the period search, allowing shorter intermediate-timescale signals to be recovered more readily. Consequently, the distribution shifts toward lower $P_{\mathrm{Rieg}}$ and shows a prominent concentration near $X_{\mathrm{R}}\simeq6$, in close agreement with the canonical solar Rieger periodicity of approximately six rotations. The window sweep also increases the number of stars satisfying the detection criteria by allowing the detrending scale to adapt to each candidate signal.

As an additional test, we applied the pipeline using a single fixed detrending-window width for all stars. Figure~\ref{fig:Prieg vs Prot color Ro no filter} shows that this produced an artificially narrow and nearly linear $P_{\mathrm{Rieg}}$--$P_{\mathrm{rot}}$ relation, with a slope determined approximately by the filter width expressed in units of $P_{\mathrm{rot}}$. This result demonstrates that a fixed filter scale can imprint a spurious relation on the detected periods. In contrast, sweeping over a range of window widths and requiring stability across the sweep substantially reduces the sensitivity of the recovered relation to the adopted detrending scale.

The lower boundary of the measured $P_{\mathrm{Rieg}}$--$P_{\mathrm{rot}}$ distribution is set by

$$
X_{\mathrm{R}}\equiv\frac{P_{\mathrm{Rieg}}}{P_{\mathrm{rot}}}=2
$$

which follows from the lower limits adopted in both the Lomb--Scargle period search and the detrending-window range, as described in Section~\ref{sec:data rieger}.

The $X_{\mathrm{R}}$ histogram in Figure~\ref{fig:xr histogram} shows that the bulk of the measured distribution lies well above the imposed lower boundary at $X_{\mathrm{R}}=2$, with no substantial accumulation of detections along the boundary. This is further supported by the density representation in panel (b) of Figure~\ref{fig:Prieg vs Prot color Ro no filter}, which shows that, across the populated range of $P_{\mathrm{rot}}$, the highest-density regions remain well separated from the lower boundary at $P_{\mathrm{Rieg}}=2P_{\mathrm{rot}}$. Together, these results indicate that the observed $P_{\mathrm{Rieg}}$--$P_{\mathrm{rot}}$ relation is unlikely to be generated by the lower period-search limit itself.

For the upper boundary, we restrict the Lomb--Scargle search to periods shorter than 300 days. The density distribution in panel (b) of Figure~\ref{fig:Prieg vs Prot color Ro no filter} shows that, throughout the populated rotation-period range, the bulk of the detections also lies well below this upper limit. In particular, the slow-rotator population is concentrated around $P_{\mathrm{Rieg}}\simeq150$ days, approximately a factor of two below the adopted cutoff, while the fast- and intermediate-rotator populations are concentrated at even shorter periods. This suggests that the dominant population structure is unlikely to be strongly affected by the upper search boundary. 

\section{Discussion}
\label{sec:discussion}

\subsection{A Rossby-Number Dynamo Reconfiguration and the Rieger-type Clock}
\label{sec:DiscussionRossbyTransition}

The Rossby-dependent transition in rotational coupling identified in Section~\ref{sec: Ro Transition} provides an important clue to the physical origin of the Rieger-type clock. We interpret $\mathrm{Ro}_{\mathrm{tr}}=0.8\pm0.1$ as potentially marking a change in the rotational--convective state of the stars and, consequently, in the organization of their magnetic dynamos.

This interpretation follows naturally from the physical significance of an order-unity Rossby number. For $\mathrm{Ro}<1$, convection evolves under a strong rotational constraint, with the Coriolis force playing an important role in organizing the flow, whereas for $\mathrm{Ro}>1$, this constraint weakens and convective inertia becomes comparatively more important. A transition between these dynamical regimes provides a natural physical basis for the observed change in $P_{\mathrm{Rieg}}$--$P_{\mathrm{rot}}$ coupling.

Independent magnetic diagnostics support the idea that such a reorganization occurs near the same Rossby-number range. Using Zeeman--Doppler imaging, \citet{See2016} found that stars at higher Rossby number are predominantly characterized by poloidal, axisymmetric large-scale magnetic fields, whereas lower-Rossby-number stars can sustain substantial toroidal and non-axisymmetric components. Independently, \citet{Lehtinen2021} identified a break in the chromospheric rotation--activity relation near $\mathrm{Ro}\sim1$ and associated it with a change between dominant dynamo configurations. Although the diagnostics and samples differ from those considered here, the proximity of these transitions suggests that the change in $P_{\mathrm{Rieg}}$--$P_{\mathrm{rot}}$ coupling may represent another manifestation of a broader Rossby-dependent reorganization of stellar magnetism.

A natural physical interpretation is that such a dynamo reconfiguration changes the behavior of the global magnetic process responsible for the Rieger-type clock. In magnetic-Rossby-wave models, the characteristic frequency is not determined by rotation alone, but also depends on quantities such as the toroidal magnetic-field strength, differential rotation, and the excited wave mode \citep{Gurgenashvili2016}. Thus, a relatively organized magnetic and shear configuration at lower Rossby number could produce the strong scaling of $P_{\mathrm{Rieg}}$ with $P_{\mathrm{rot}}$ observed in the intermediate regime. As $\mathrm{Ro}$ approaches unity, changes in magnetic topology, shear, or the dominant global mode could introduce additional control of $P_{\mathrm{Rieg}}$ by the broader magnetic state, thereby weakening its direct dependence on rotation.

A corresponding change in differential rotation is also expected theoretically. Global convection simulations show a transition between rotation-dominated, solar-like differential rotation and a more buoyancy-dominated, anti-solar state when rotational constraint becomes sufficiently weak \citep{Gastine2014}. The Rossby numbers defined in such simulations are not directly equivalent to the observational Rossby number adopted here, so the numerical transition values should not be compared literally. Nevertheless, these calculations demonstrate that the order-unity Rossby regime is dynamically important, because the balance between Coriolis forces and convective inertia can reorganize angular-momentum transport and large-scale shear. Such a change provides a plausible route through which the dynamo and the spectrum of magnetically modified global modes could change together.

An alternative interpretation is that the transition reflects a broader reorganization of the stellar magnetic state, potentially associated with angular-momentum redistribution between the radiative interior and convective envelope. \citet{Mathur2025} identified a non-monotonic feature in the $S_{\mathrm{ph}}$--Rossby-number relation of G and K dwarfs, consisting of an activity minimum near $\mathrm{Ro}/\mathrm{Ro}_{\odot}\simeq0.3$ followed by an increase to a local peak near $\mathrm{Ro}/\mathrm{Ro}_{\odot}\simeq0.4$. Outside this interval, the magnetic activity follows the more usual decrease with increasing Rossby number, whereas between the dip and the peak the trend is temporarily reversed. They associated the dip with the intermediate rotation-period gap and suggested that it may mark the onset of angular-momentum redistribution between the radiative interior and convective envelope, although it remains unclear whether the associated reconfiguration extends throughout the full dip-to-peak interval.

Our transition occurs at $\mathrm{Ro}_{\mathrm{tr}}=0.8\pm0.1$ ($\mathrm{Ro}\simeq0.7$--$0.9$). When expressed on the solar-normalized Rossby scale used by \citet{Mathur2025}, this corresponds approximately to $\mathrm{Ro}/\mathrm{Ro}_{\odot}\simeq0.3$--$0.4$, overlapping their entire dip-to-peak region. Given the uncertainty in our transition location, we cannot determine whether the change in $P_{\mathrm{Rieg}}$--$P_{\mathrm{rot}}$ coupling corresponds specifically to the activity minimum, the subsequent rise, or the peak itself. Rather, the important correspondence is that \citet{Mathur2025} find anomalous magnetic-activity evolution throughout approximately the same Rossby-number interval in which we find that the behavior of the Rieger-type clock begins to change. The two diagnostics may therefore be tracing different manifestations of the same broader reorganization of the rotational--convective and magnetic state.

The presence of Rieger-type variability in fully convective stars (Section~\ref{sec:SlowInterFastRegimes}) provides an important constraint on the core--envelope interpretation. If the underlying mechanism is shared by partly and fully convective stars, neither core--envelope angular-momentum redistribution nor the confinement of global modes to a solar-like tachocline can provide a universal explanation for the Rieger-type phenomenon. Its persistence across this structural boundary instead favors a more general dynamical process that can operate in different stellar interiors, such as global magnetically modified Rossby-type or related modes, although the detailed mode structure need not be identical in both populations.

A complementary case is provided by chemically peculiar Ap/Bp stars, in which the presence of Rieger-like variability suggests that intermediate periodicities may also occur in stars dominated by strong, stable fossil magnetic fields. Although the small sample prevents us from constraining the underlying mechanism, these objects illustrate the potentially broad range of magnetic configurations in which Rieger-like variability may occur.

Taken together, we interpret the transition near $\mathrm{Ro}\sim0.8$ as evidence for a reconfiguration of the stellar dynamo, in which the changing balance between rotation and convection alters the organization of the magnetic field and the behavior of associated global magnetic or Rossby-type modes. In this picture, the Rieger-type period is not controlled by rotation alone, but by a broader magnetic and dynamical state that also depends on convection, differential rotation, and field geometry. As the dynamo reorganizes near order-unity Rossby number, the dominant global-mode behavior may change, producing the observed transition from a strongly coupled $P_{\mathrm{Rieg}}$--$P_{\mathrm{rot}}$ relation in the intermediate regime to the much weaker rotational dependence of the slow regime. Core--envelope angular-momentum redistribution may contribute to this reconfiguration in stars possessing a radiative interior, but the fully convective detections argue against it as the universal mechanism underlying the Rieger-type clock.

\subsection{Tidally Synchronized Binaries as a Test of Surface--Interior Coupling}
\label{sec:DiscussionBinaries}

The tidally synchronized binaries identified in Section~\ref{sec:Binaries} provide a natural experiment for investigating the response of the Rieger-type clock to an externally imposed change in surface rotation. Despite their rapid surface rotation, these systems retain $P_{\mathrm{Rieg}}$ values and stellar properties, including effective temperature, mass, radius, metallicity, and surface gravity, characteristic of the slow-rotator population.

This behavior suggests an important distinction between the observed surface rotation period and the rotational state relevant to the Rieger-type phenomenon. In normal single stars, the measured $P_{\mathrm{rot}}$ is expected to evolve together with the internal rotational state and can therefore serve as an observable proxy for it. Tidal synchronization can break this correspondence by forcing the outer layers toward the orbital period while the deeper rotational and magnetic configuration need not respond in the same way. Consequently, the extremely low observed Rossby numbers of these systems reflect their tidally shortened surface rotation and need not characterize their deeper rotational state. The persistence of slow-like $P_{\mathrm{Rieg}}$ is therefore consistent with the synchronized systems retaining the magnetic--dynamical state of the slow regime despite their rapid surface rotation.

The intermediate population can be understood within the same framework. Its strong $P_{\mathrm{Rieg}}$--$P_{\mathrm{rot}}$ dependence does not necessarily imply that the Rieger-type timescale is set directly by photospheric rotation; rather, in normal stars, the observed $P_{\mathrm{rot}}$ may closely track the internal rotational state. The stronger coupling in the intermediate regime may therefore reflect a different dynamical regime or mode family in which internal rotation plays a dominant role in setting $P_{\mathrm{Rieg}}$. In this picture, the slow-to-intermediate transition is not simply a change in the slope of a universal surface-rotation relation. The synchronized binaries suggest that shortening the surface rotation period alone is insufficient to trigger this transition; a corresponding change in the internal rotation--convection state may also be required.

This interpretation further weakens a simple core--envelope interaction model in which $P_{\mathrm{Rieg}}$ is set primarily by the rotational contrast between the two regions, since tidal acceleration of the envelope should substantially modify that contrast. The persistence of slow-like $P_{\mathrm{Rieg}}$ despite this perturbation suggests that core--envelope shear alone is unlikely to be the dominant quantity setting the Rieger-type timescale, although the internal rotation profiles of these systems are not directly measured.

Beyond their implications for surface--interior coupling, the clear separation between tidally synchronized binaries and intrinsically rapid rotators in the $P_{\mathrm{Rieg}}$--$P_{\mathrm{rot}}$ plane suggests that combining these two periods provides a purely photometric means of distinguishing between the two populations, without requiring independent age estimates or orbital measurements. This offers a physically motivated alternative to previously employed statistical and multidimensional classification approaches. Given the high incidence of tertiary companions among short-period binaries, this approach may also provide an indirect means of identifying candidate hierarchical triple systems.

\subsection{Connection between the Rieger-type Clock and Surface Magnetic Coherence}
\label{sec:DiscussionSpotLifetime}

The correspondence between $P_{\mathrm{Rieg}}$ and the spot-pattern coherence time identified in Section~\ref{sec:spot_lifetime} provides a possible link between the Rieger-type clock and the evolution of surface magnetic structures.

Within the dynamo-reconfiguration picture developed above, this relation may represent the surface manifestation of a deeper magnetic process. If the Rieger-type clock is associated with a global magnetic or Rossby-type mode, recurrent modulation of flux emergence or large-scale magnetic organization could naturally influence the formation, replacement, and reorganization of active regions at the surface. In this case, the similar values of $P_{\mathrm{Rieg}}$ and $L_{\mathrm{spot}}$ would reflect a causal connection between the global magnetic clock and the lifetime of the observable surface pattern rather than two independent timescales.

A simpler possibility is that the near-unity relation partly reflects the fact that both $P_{\mathrm{Rieg}}$ and $L_{\mathrm{spot}}$ are extracted from the same photometric light curves using related diagnostics of rotational variability. Evolving spots, active longitudes, or differential rotation could therefore influence both measurements, so the correspondence alone does not establish a distinct magnetic clock.

Forward modeling of spot evolution could help distinguish between these possibilities.

\section{Summary and Conclusions}
\label{sec:conclusions}

In this work, we developed a dedicated procedure to measure Rieger-type periodicities using \emph{Kepler} photometry and rotation periods from four independent catalogs. We measured $P_{\mathrm{Rieg}}$ for 15,319 unique stars, constructing, to our knowledge, the largest catalog of individually measured stellar Rieger-type periods to date.

The $P_{\mathrm{Rieg}}$--$P_{\mathrm{rot}}$ plane reveals four distinct stellar populations: slow rotators, intermediate rotators, a fast-low-$P_{\mathrm{Rieg}}$ population associated with $\gamma$~Doradus stars, and a fast-high-$P_{\mathrm{Rieg}}$ population dominated by tidally synchronized binaries. The intermediate population exhibits strong rotational coupling, whereas the slow and hot fast populations show only weak dependence on the measured photometric period, with characteristic Rieger-type periods of approximately 150 and 18~d, respectively. The characteristic period of the slow population and the prominent concentration near $X_{\mathrm{R}}=P_{\mathrm{Rieg}}/P_{\mathrm{rot}}\simeq6$ are both consistent with the canonical solar Rieger periodicity.

For the calibrated late-type sample, the transition between the strongly coupled intermediate regime and the weakly coupled slow regime occurs near $\mathrm{Ro}_{\mathrm{tr}}=0.8\pm0.1$. The Rossby-number dependence of this transition and its proximity to independently reported changes in stellar magnetic topology and activity \citep{See2016,Mathur2025} support its interpretation as a stellar dynamo reorganization.

The tidally synchronized binaries provide complementary evidence for this interpretation. Despite their anomalously rapid surface rotation, these systems retain stellar properties and Rieger-type periods characteristic of ordinary slow rotators, suggesting that the Rieger-type clock is sensitive to the underlying magnetic--dynamical state rather than to surface rotation alone. Their distinct location in the $P_{\mathrm{Rieg}}$--$P_{\mathrm{rot}}$ plane also offers a potential purely photometric diagnostic of tidally synchronized binary systems.

The detection of Rieger-type variability in fully convective stars challenges mechanisms requiring a solar-like tachocline, while Rieger-like variability in chemically peculiar Ap/Bp stars suggests that intermediate periodicities may occur under diverse magnetic configurations. The correspondence between $P_{\mathrm{Rieg}}$ and the spot-pattern coherence time, further suggests a connection between the Rieger-type clock and surface magnetic evolution.

Taken together, our findings suggest that Rieger-type periodicity may trace a global magnetic process governed by the underlying magnetic--dynamical state of the star, whose coupling to rotation changes across a Rossby-dependent dynamo transition. Global magnetically modified Rossby-type modes remain a plausible mechanism, although the specific physical origin of the Rieger-type clock remains unidentified. These results establish Rieger-type variability as a promising population-level diagnostic of stellar magnetic dynamics.

Future work will investigate the temporal modulation of $P_{\mathrm{Rieg}}$, its relation to the long-term activity-cycle period $P_{\mathrm{cyc}}$, and the behavior of secondary Rieger-type modes. A key objective will be to determine whether these periodicities can provide empirical constraints on stellar magnetic-field strength, the mixing-length parameter $\alpha_{\mathrm{MLT}}$, stellar age, and other properties linked to dynamo evolution. Combining the photometric identification of synchronized binaries with existing multiplicity catalogs may also enable the identification and statistical characterization of hierarchical triple systems.

\section*{Acknowledgments}

The authors acknowledge the use of OpenAI's ChatGPT and Anthropic's Claude exclusively for assistance with manuscript editing, language refinement, and organization. These tools were not used for data processing, calculations, statistical analysis, or the generation of scientific results. All scientific analyses, interpretations, and conclusions were developed and independently verified by the authors. Furthermore, all authors carefully reviewed the entire manuscript multiple times, including its scientific content, numerical values, and references, to ensure accuracy and identify any potential AI-generated errors or hallucinations. The authors take full responsibility for the accuracy, integrity, and content of this work.

\section*{Data Availability}

The \emph{Kepler} photometric data used in this work are publicly available through the Mikulski Archive for Space Telescopes (MAST). The stellar rotation periods and stellar properties were obtained from the publicly available catalogs cited in this work. Two catalogs will be made publicly available upon publication: the full catalog of measured Rieger-type periodicities and the filtered catalog of 15,319 unique stars satisfying the selection criteria adopted in this work.

\bibliographystyle{apsrev4-1}
\bibliography{main}

\end{document}